%% file: Main_Supplementary_Merged.tex
\documentclass[aps,prl,reprint,groupedaddress,twocolumn]{revtex4-2}
\usepackage{graphicx}
\usepackage{xcolor}
\usepackage{amsmath}
\usepackage{amssymb}
\usepackage{bm}
\usepackage{ulem}
\usepackage{svg}
\DeclareMathOperator*{\argmax}{arg\,max}

\begin{document}


\title{Quantum Conformance Test}
\author{Giuseppe Ortolano$^{1,2}$}
\author{Pauline Boucher$^1$}
\author{Ivo Pietro Degiovanni$^1$}
\author{Elena Losero$^{1,3}$}
\author{Marco Genovese$^1$}
\author{Ivano Ruo-Berchera$^1$}
\affiliation{$^1$Quantum metrology and nano technologies division,  INRiM,  Strada delle Cacce 91, 10153 Torino, Italy}
\affiliation{$^2$DISAT, Politecnico di Torino, Corso Duca degli Abruzzi 24,
10129 Torino, Italy}
\affiliation{$^3$ STI-SB, Ecole Polytechnique Fédérale de Lausanne, Lausanne, 1015, Switzerland}

\begin{abstract}
We introduce a protocol addressing the \textit{conformance test} problem, which consists in determining whether a process under test conforms to a reference one. We consider a process to be characterized by the set of end-product it produces, which is generated according to a given probability distribution. We formulate the problem in the context of hypothesis testing and consider the specific case in which the objects can be modeled as pure loss channels. We demonstrate theoretically that a simple quantum strategy, using readily available resources and measurement schemes in the form of two-mode squeezed vacuum and photon-counting, can outperform any classical strategy. We experimentally implement this protocol, exploiting optical twin beams, validating our theoretical results, and demonstrating that, in this task, there is a quantum advantage in a realistic setting.
\end{abstract}


\maketitle

\section{Introduction}
Significant progress has been made in recent years in the field of quantum sensing \cite{Pirandola_2018,Degen_2017} both for continuous parameter estimation \cite{Genovese_2016,Berchera_2019, Polino20}, and for discrimination tasks in the case of discrete variables \cite{Helstrom_1976, Chefles_1998, Chefles_2000}. The use of quantum states and resources brings an advantage that has been demonstrated in various specific tasks: both phase \cite{Aasi_2013,Berchera_2013,Schafermeier_2018,Ortolano_2019,Pradyumna_2020} and loss \cite{Tapster_1991,Monras_2007,Adesso_2009,Matthews_2016,Moreau_2017,Losero_2018, Avella_2016} estimation, quantum imaging \cite{Brida_2010,Samantaray_2017,Sabines-Chesterking_2019}, and discrimination protocols such as  target detection \cite{Lloyd_2008,Tan_2008,Lopaeva_2013,Zhang_2015}  and quantum reading \cite{Pirandola_2011,Ortolano_2021}. 

In this class of problems, a parameter of interest is encoded in some quantum states, or channels, and the values it can take can belong to either discrete or continuous sets. The estimation of this parameter requires the choice of a probe, as well as a measurement scheme, i.e. a measurement strategy. Here, we consider a significant discrimination problem that can find various important applications, the quantum conformance test. In this problem, one wants to assess whether a process is conform to a reference, or if it is defective. In the general case, the process is characterized by a physical parameter, distributed according to some continuous probability density distribution. We experimentally have access to the end-products of this process, which we can perform measurements on. Accordingly, the measurement outcomes belong to a continuous set of possible values. However, the expected output of the procedure should be binary: conform or not conform. Such conformity tests appear frequently in many applications \cite{Pendrill_2014}, one example being product safety testing.

Under energy constraints for the probe, quantum mechanical fluctuations set lower bounds to the probability of error that depend on the decision strategy. It is important to investigate whether, and to what extent, the use of certain quantum resources can reduce the probability of error below what is possible in the classical domain.

In this work, we introduce a formal description of the conformance test problem in the context of quantum information \cite{Nielsen_2011} and consider the paradigmatic example of bosonic pure loss channels probed with light. We demonstrate that, under the same energy constraints, i.e. fixing the photon number, a quantum strategy making use of entangled photons as probe states, and photon-counting (PC) as of measurement strategy, can deliver better results than any classical strategy. Finally, we present an experimental optical implementation of our proposed quantum protocol, which validates our theoretical model and shows that a genuine quantum advantage persists even in presence of experimental inefficiencies.

\subsection{Quantum Conformance test}
\begin{figure*}
\vspace{0.10cm}
\includegraphics[width=2\columnwidth]{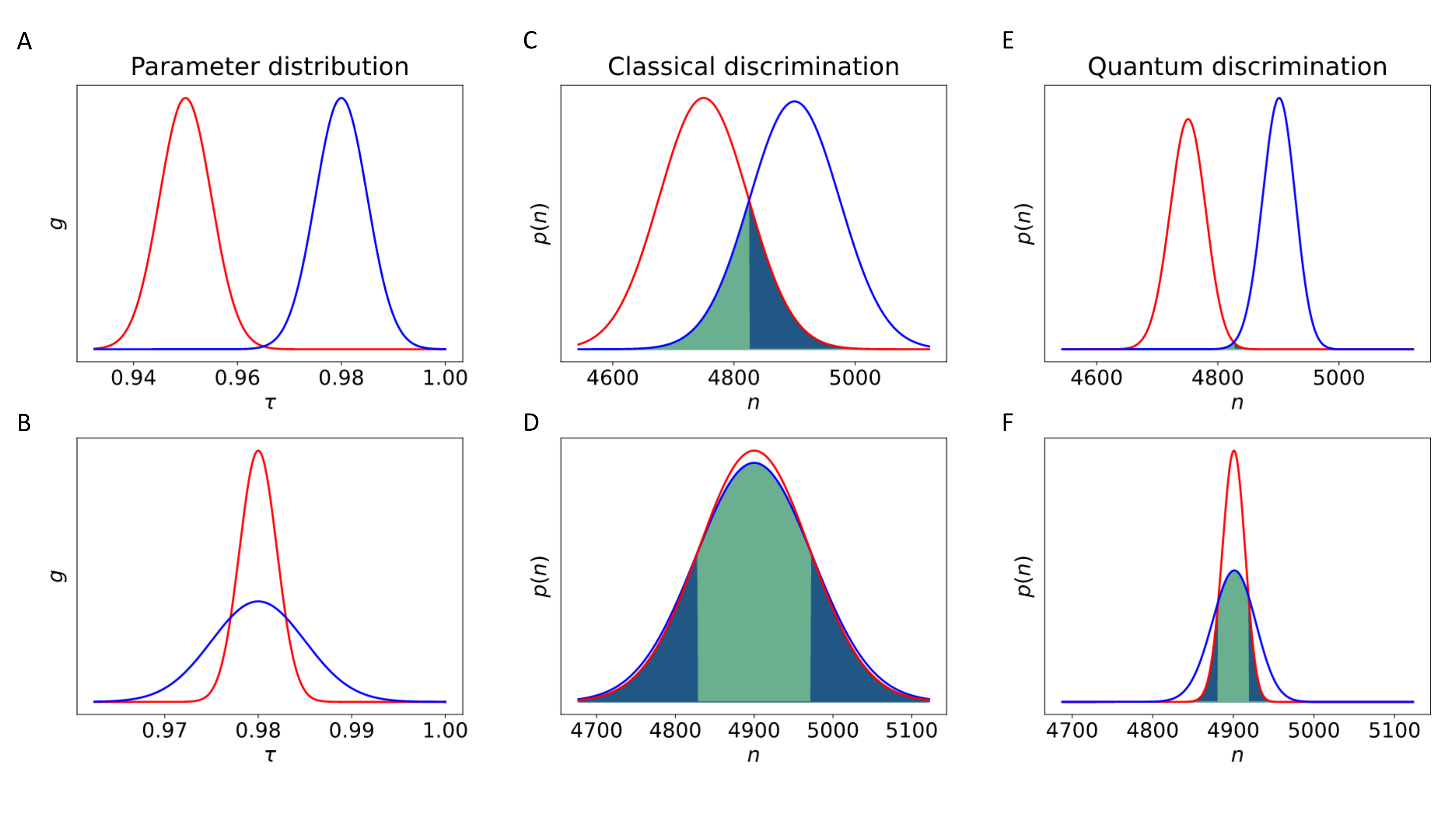}
\caption{\label{fig:com}\emph{Examples of error probabilities for the conformance test.} In panels \textbf{A-B}, the distributions of the reference process (red), $\mathcal{P}_0$, and defective one (blue), $\mathcal{P}_1$, are shown: the two rows present opposite and archetypal situations. The graphics (C-D-E-F) are obtained considering a pure loss channel of parameter $\theta=\tau$ as the SUT, and a PC measurement. In panels \textbf{C-D}, the resulting photon number distribution $p(n)$, in the case where a classical state is used as a probe, are shown. The overlaps between the two distributions, highlighted in green and blue, the two colors distinguishing the cases where the reference or the defective process are most likely, are visualizations of the conditional error probabilities, at a given value of $\tau$, while their weighted sum gives an appreciation of the total error probability. Panels \textbf{E-F} display the case in which quantum probes are used: quantum correlations enhance the performance of the discrimination task.}
\end{figure*}

The quantum conformance test (QCT) can be modeled as follows. We define the binary random variable $x\in\{0,1\}$, which corresponds to the process from which the physical system under test (SUT) was generated, $0$ being the reference and $1$ the defective one. We consider the monitoring of a physical process $\mathcal{P}_x$, producing a quantum object (the SUT), $\mathcal{E}_{\theta}$, which depends on a parameter $\theta$. The process $\mathcal{P}_x$ can be described by the ensemble $\{g_x(\theta), \mathcal{E}_{\theta}\}$: the parameter $\theta$ is extracted from a set $\mathcal{A}$, according to the probability distribution $g_x(\theta)$, and it defines the physical object $\mathcal{E}_{\theta}$. The set $\mathcal{A}$ can either be discrete or continuous. The conformance test consists in ruling whether an unknown process is conform to a ``reference" process, $\mathcal{P}_0$, or if it should be labeled ``defective", $\mathcal{P}_1$, using measurements on the set of objects it produces $\{\mathcal{E}_{\theta}\}$.

A QCT is performed using a probe in a generic quantum state $\rho$. After the probe has interacted with the SUT, the final state is measured by a positive-operator-valued measure (POVM) $\Pi$ and, after classical post-processing of the measurement results, the outcome of the procedure is a final guess on the nature of the production process, expressed by the binary variable $y\in\{0,1\}$.

The test is successful when $y=x$, i.e. when the guess is correct. On the other hand, if $y \neq x$, the test fails. Two cases can be distinguished:
\begin{itemize}
\item
\textit{False negative}. In this scenario, a SUT produced by a conform process ($x=0$) is labeled as defective ($y=1$). In an industrial context, this kind of outcome can be seen as an economic loss for a manufacturer, as a conform process is considered defective. We will denote the probability of false negatives as $p_{10}$.
\item
\textit{False positive}. A SUT produced by a defective process ($x=1$) is labeled as conform ($y=0$). This outcome represents a risk since possibly unsafe products are released. The false-positive probability will be referred to as $p_{01}$.
\end{itemize}
The analysis of false positives and negatives plays a central part in conformity testing, and the specific choice, if required, of the tolerance on either one of those errors vastly depends on the situation considered. In a general scenario a significant figure of merit -- when the energy of the probe is considered a limited resource and, therefore one compares the results at fixed energy --- that can be used to assess the effectiveness of the test is the \textit{total probability of error}, defined as:
\begin{equation}
\label{eq:1}
p_{err}=\frac{1}{2}\big(p_{01}+p_{10}\big)
\end{equation}
In Fig.(\ref{fig:com}), we present some examples of error probabilities for the conformance test. In panels \textbf{A-B}, two possible scenarios for the conform and defective processes are shown. In panel \textbf{A}, we consider two distributions whose overlap is negligible, whereas, in panel \textbf{B}, we consider significantly overlapping distributions. In panels \textbf{C-D}, we show the probability distributions for the parameter, convoluted with the noise emerging from both the measurement process and the probe state. From these distributions, the outcome $y$ must be decided. One can observe that the noise present in the state which probes the SUT translates into an error in the discrimination.

In the subsequent parts of this work, we focus on a specific class of SUTs: the bosonic pure loss channels. We show how, in this case, quantum resources can be used to significantly mitigate the spurious effect of the noise, and can improve the discrimination performance -- evaluated in terms of total probability of error -- over \textit{any classical strategy}. Subsequently, we analyze the case in which a constraint is imposed either on the false positives or the false negatives probabilities.
  
We note that the quantum reading problem, originally proposed in \cite{Pirandola_2011} and experimentally tested in \cite{Ortolano_2021}, can be considered to be a special case of the conformance test (i.e. in the case where $g_0$ and $g_1$ are both Dirac delta functions).

\section{Results}
\subsection{Pure loss channels}
\begin{figure}[h]
\centering
\def\svgwidth{\columnwidth}
    \resizebox{0.45\textwidth}{!}{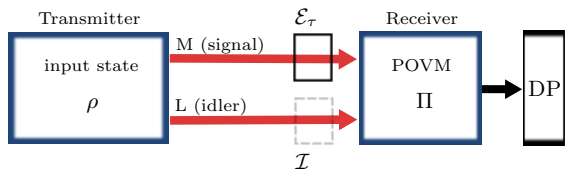}
	\caption{\emph{Quantum conformance test:} A probe $\rho$ sends $M$ signal modes through the loss channel $\mathcal{E}_{\tau}$, while $L$ idler modes directly reach the receiver -- $\mathcal{I}$ represents the identity operator. A positive-operator-valued measure $\Pi$ is applied to the output state. Using the result of this measurement and data processing (DP), a decision $y$ is taken: the process generating $\mathcal{E}_{\tau}$ is identified as conform ($y=0$) or defective ($y=1$).}
	\label{fig:exp1}
\end{figure}

A bosonic loss channel $\mathcal{E}_{\tau}$ is characterized by its transmittance $\tau \in \left[0,1\right]$. The input-output relation for such a channel is $\hat{a} \rightarrow \sqrt{\tau} \, \hat{a} + i\sqrt{1-\tau}\,\hat{v}$, with $\hat{a}$ the annihilation operator of the input mode \cite{Meda_2017}. The annihilation operator of the mode on the second port, $\hat{v}$, acts on the vacuum, as we are considering pure losses. The configuration for a conformance test over a pure loss channel is described in Fig.(\ref{fig:exp1}). A transmitter irradiates an optical probe state $\rho$ on the SUT. In a general case, the state is bipartite, having $M$ signal modes and $L$ idler ones. The state is measured at the receiver by a joint measurement, and its outputs are processed to obtain the final outcome.

After the interaction with the SUT, an input state $\rho$ will be mapped into either $\rho_0$ or $\rho_1$, with
\begin{align}
\rho_0 &= \mathbb{E}_{\mathcal{P}_0}\left[\left(\mathcal{E}_{\tau} \otimes \mathcal{I}\right)\rho\right] \rightarrow \text{reference process} \notag \\
\rho_1 &= \mathbb{E}_{\mathcal{P}_1}\left[\left(\mathcal{E}_{\tau} \otimes \mathcal{I}\right)\rho\right]  \rightarrow \text{defective process}
\label{eq:rho01}
\end{align}
and where $\mathbb{E}_\mathcal{P}[\cdot]$ represents the expectation value over the ensemble $\mathcal{P}$.

In general, the output states of the reference and defective processes, $\rho_0$ or $\rho_1$, will overlap. Accordingly, the discrimination between the processes will be affected by an error probability $p_{err}^{\rho, \Pi}(\mathcal{P}_0,\mathcal{P}_1)$ which is a function of the processes considered, the input state and the POVM -- as well as the decision procedure applied to the measurement result.
The optimization of the QCT protocol is achieved by a minimization of $p_{err}^{\rho, \Pi}(\mathcal{P}_0,\mathcal{P}_1)$ over all possible input states $\rho$, and POVMs $\Pi$:
\begin{equation}
p_{err}(\mathcal{P}_0,\mathcal{P}_1) = \min_{\Pi}\bigg[\min_\rho\big[p_{err}^{\rho,\Pi}(\mathcal{P}_0,\mathcal{P}_1)\big]\bigg]
\end{equation}
Without constraints on the energy, a trivial strategy is to let the energy of the probe system go to infinity, which, given a suitable measurement, would nullify the quantum noise and lead to the minimum possible $p_{err}$ permitted by the problem, i.e. the overlap between the two initial distributions $g_0$ and $g_1$ (see Fig.(\ref{fig:com})). Nevertheless, in several significant cases, one cannot dispose of, or use, states of arbitrarily large energy (e.g. in order not to damage the SUT). In the case we are considering, we fix the total energy of the probe, and the optimization problem is not easily solved directly. Following the approaches used, for example in \cite{Pirandola_2011, Pirandola_2011a}, the minimum probability of error that can be achieved with any classical transmitter can be derived. Then, if a class of quantum transmitters whose probability of error is lower than that limit, can be identified, a quantum advantage is demonstrated. This is the approach we adopt.

\subsection{Classical limit}

First, we turn our attention to the derivation of the minimum error probability which can be achieved using classical input states. In the context of quantum optics, classical states can be defined as the states having a positive P-representation \cite{Mandel_1995}. A generic classical bipartite state can be represented as:
\begin{equation}
\rho^{cla} = \int d^{\text{\tiny$2M$}}\bm{\alpha}\, d^{\text{\tiny$2L$}} \bm{\beta} \, P\left(\bm{\alpha}, \bm{\beta} \right)|\bm{\alpha}\rangle \langle \bm{\alpha}| \otimes |\bm{\beta}\rangle \langle \bm{\beta}|
\label{eq:classicalstate}
\end{equation}
with $|\bm{\alpha}\rangle$ and $|\bm{\beta}\rangle$ respectively $M$- and $L$-mode coherent states (corresponding to the signal and idler channels), and $P\left(\bm{\alpha}, \bm{\beta} \right) \geq 0$ a probability density. The energy constraint for the signal system is expressed, in terms of mean signal photon number $\bar{n}_S$, as:
\begin{equation}
\int d^{\text{\tiny$2M$}} \bm{\alpha}\,  d^{\text{\tiny$2L$}} \bm{\beta} \, P\left(\bm{\alpha},\bm{\beta} \right)|\bm{\alpha}|^2 = \bar{n}_S
\end{equation}
While the choice of imposing the energy constraint only on the signal system is arbitrary -- the alternative being for example a constraint on the total energy -- it is a natural choice for real applications, where the energy irradiated over the SUT should be limited. 

Given an input state $\rho^{cla}$, the output states, after interaction with the SUT, are $\rho_0^{cla}$ and $\rho_1^{cla}$, calculated according to Eq.(\ref{eq:rho01}). In this context, the minimum probability of error for the QCT protocol with classical states is equal to the minimum probability of error in the discrimination of $\rho_0^{cla}$ and $\rho_1^{cla}$. The best performance in this task is achieved by using a POVM measurement, which assumes the Helstrom projectors as elements \cite{Helstrom_1976}. This optimal discrimination procedure yields a probability of error given by:
\begin{equation}
p_{err} = \frac{1}{2}\left(1 - D\left(\rho_0, \rho_1\right)\right)
\end{equation}
where $D\left(\rho_0, \rho_1\right) = ||\rho_0 - \rho_1||/2$ is the trace distance with $||\rho||=\sqrt{\rho^{\dagger}\rho}$. A lower limit for $p^{cla}_{err}$ can be found by upper bounding $D\left(\rho_0, \rho_1\right)$. By exploiting the convexity of the trace distance (see the Supplementary Materials for details), the minimum error probability for classical states in the QCT protocol is bounded by:
\begin{equation}
\mathcal{C}:=p^{cla}_{err} \geq \frac{ 1- \mathbb{E}_{\mathcal{P}_0}\left[\mathbb{E}_{\mathcal{P}_1}\left[\sqrt{1-e^{-n_S\left(\sqrt{\tau_0} - \sqrt{\tau_1 }\right)^2}}\right]\right]}{2}
\label{eq:classicalbound}
\end{equation}
The quantity $\mathcal{C}$ establishes a lower bound for the discrimination error probability when considering classical resources and an optimal measurement strategy. We note that this bound is not tight, which means it may not be reached by any classical receiver.
 
\subsection{Quantum strategy}
In the following we analyze the particular strategy, involving quantum states, that is able to surpass the best classical performance $\mathcal{C}$. It uses a transmitter $\rho$ constituted of $K$ replicas of a two-mode squeezed vacuum (TMSV) state and a PC receiver \cite{Ortolano_2021}, whose output is processed by a maximum likelihood decision.

The TMSV state \cite{Weedbrook_2012} admits the following expression in the photon number basis: $|\psi\rangle=\sum_{n=0}^{\infty} \sqrt{P_{\bar{n}}(n)} |n\rangle_S|n\rangle_I$ with $P_{\bar{n}}(n)$ a thermal distribution, $\bar{n}$ the mean photon number and $|n\rangle_i$ the $n$-photon state in the $i=S$ signal, or $i=I$ idler, mode. TMSV states can easily be produced experimentally by parametric down-conversion \cite{Bondani_2007,Perina_2007,Chekhova_2018} or four wave mixing \cite{Glorieux_2011,Pooser_2016,Wu_2019}. A multimode TMSV state is a tensor product of $K$ TMSV states $\otimes^K|\psi\rangle$. It admits a multi-thermal distribution for the total photon number both in the signal and idler modes, $n_{i} = \sum_{k=1}^K n_i^{(k)}$, denoted as $P_{\bar{n},K}(n_i)$ (the same notation $\bar{n}$ is used for the mean photon number) and preserves perfect photon number correlation between the channels i.e. $P_{\bar{n},K}\left(n_S, n_I\right) = P_{\bar{n},K}\left(n_I\right)\delta_{n_S,n_I}$. The result of the PC measurement is the classical random variable $\bm{n}=\left(n_S, n_I\right)$, whose distribution $p\left(\bm{n}|\mathcal{P}_x\right) = \langle n_S, n_I |\rho_x|n_S, n_I\rangle$, is conditioned on the nature $x$ of the process, with $\rho_x$ defined in equation \ref{eq:rho01}. Using Bayes theorem and assuming the defective and reference processes to be equiprobable, $p\left(\mathcal{P}_x = \mathcal{P}_0\right)=p\left(\mathcal{P}_x =\mathcal{P}_1\right)=1/2$, we can write the \textit{a posteriori} probability $p\left(\mathcal{P}_x|\bm{n}\right)$ as:
\begin{equation}
p\left(\mathcal{P}_x|\bm{n}\right) = \frac{p\left(\bm{n}|\mathcal{P}_x\right)p\left(\mathcal{P}_x\right)}{p\left(\bm{n}\right)} = \frac{p\left(\bm{n}|\mathcal{P}_x\right)}{p\left(\bm{n}|\mathcal{P}_0\right) + p\left(\bm{n}|\mathcal{P}_1\right)}
\end{equation}
Let's first consider the action of a pure loss channel on the photon number distribution of a state. The distribution $p\left(\bm{n}|\tau\right)$ after the interaction is the composition of the joint distribution $P_{\bar{n},K}\left(n_S, n_I\right)$  with a binomial distribution:
\begin{equation}
p\left(\bm{n}|\tau\right) = \sum_{m=n_S}^{\infty}P_{\bar{n},K}\left(m, n_I\right) \mathcal{B}\left(n_S|m, \tau\right).
\end{equation}
since the $m$ signal photons can be seen as undergoing a Bernoulli trial each, with probability of success $\tau$, resulting in the binomial distribution $\mathcal{B}\left(n_S|m, \tau\right)$. Using the linearity of quantum operations, we can evaluate the effect of a loss channel with transmittance $\tau$, $p\left(\bm{n}|\mathcal{P}_x\right)$, as:
\begin{equation}
p\left(\bm{n}|\mathcal{P}_x\right) = \int_{\mathcal{A}_x} p\left(\bm{n}|\tau\right) g_{x}\left(\tau\right)d\tau.
\end{equation}
After the measurement, the decision is made by choosing the outcome which maximizes the conditional probability: $y$ is chosen such that $y=\argmax_{x\in\{0,1\}}p\left(\mathcal{P}_x|\bm{n}\right)$. We note that this condition for the choice of $y$ is equivalent, due the constant prior, to a maximum likelihood decision, i.e. choosing $y$ such that $p\left(\bm{n}|\mathcal{P}_y\right)\geq p\left(\bm{n}|\mathcal{P}_{	1- y}\right)$. The probability of error for the quantum strategy, $\mathcal{Q}$, is given by:
\begin{align}
\mathcal{Q}:=p_{err}^{QCT} &= \sum_{\bm{n}} \min_{x}p\left(\mathcal{P}_x |\bm{n}\right)p\left(\bm{n}\right) \notag \\
&= \frac{1}{2}\sum_{\bm{n}} \min_{x}p\left(\bm{n}|\mathcal{P}_x \right) \label{qperr}
\end{align}

\subsection{Classical states and photon counting}
We have derived a limit $\mathcal{C}$ on the performance that can be achieved with an optimal classical strategy, i.e. optimal classical input states and an unspecified optimal receiver. We also defined the performance,$\mathcal{Q}$, of a quantum strategy using TMSV and PC. In this section, we consider the case in which classical states are paired with a PC receiver. Indeed, the bound found in Eq.(\ref{eq:classicalbound}) is not tight, meaning that it may not be possible to reach it. Moreover, in the case where a POVM could be found that saturates the bound, its implementation may be of difficult practical realization. The analysis of the best classical performance achievable with the PC receiver will give a second classical benchmark, whose performance can be experimentally validated.

The analysis is analogous to the one performed for the quantum strategy. However, since the input states considered are limited to classical ones, the use of idler modes cannot improve the performance. Classical states are statistical mixtures of coherent states, that are Poisson distributed in the photon number: their variance is lower bounded by the Poisson one \cite{Mandel_1995}. In this scenario, the best performance is achieved using signal states having a Poisson photon number distribution. Indeed, the error probability $p_{err}$ is proportional to the overlap of the measurement outcomes, as shown in Fig.(\ref{fig:com}), which, in the case of a PC measurement, are the photon number distributions. Using a state with a narrower photon number distribution will lead to better discrimination performances. 
We denote the best performance which can be achieved using classical states and PC as $\mathcal{C}^{pc}$, and write it in the form:
\begin{equation}\label{eq:classicalboundpc}
\mathcal{C}^{pc}:=p^{cla,pc}_{err}=\frac{1}{2}\Big(1-q_p\Big)
\end{equation}
where $0\leq q_p \leq 1$. The form of the function $q_p$ depends on the distributions of the considered processes, $g_0$ and $g_1$. In the following, we report on the case where both the reference and defective process have Gaussian distributions $G_{\bar{\tau}, \sigma}$ (with $\bar{\tau}$ and $\sigma^2$ the mean and the variance). The distributions we consider are thus $g_0=G_{\bar{\tau}_0,\sigma_0}$ and $g_1=G_{\bar{\tau}_1,\sigma_1}$ for the reference and defective processes respectively. For more general solutions, and a more in depth analysis, we refer the reader to the Supplementary Materials. 

As pointed out in previous sections, the probability of error depends on the overlap of the two measurement distributions $p\left(\mathcal{P}_0|n_S\right)$ and $p\left(\mathcal{P}_1 |n_S\right)$. Under the assumption of Gaussian distributions for both processes, and of a large photon number in the initial probe state, $\bar{n}_s \gg 1$, the photon number distribution at the outcome can be well approximated by a Gaussian: $p\left(\mathcal{P}_x|n\right) \approx G_{\bar{\tau}_x \bar{n}_s,\sigma_{G^{(x)}}}$,
with $(\sigma_G^{(x)})^{2}=\bar{n}_s\bar{\tau}_x +\bar{n}_s ^2 \sigma_x^2$ (details can be found in Supplementary Materials). An expression for the overlap can be found, after the determination of the solutions of the equality $p\left(\mathcal{P}_0|n\right)=p\left(\mathcal{P}_1 |n\right)$. In general, this equation admits two solutions in the case where the distributions are Gaussian. If we can assume the variances $\sigma_G^{(x)}$ to be of the same order, and $\bar{\tau}_0 < \bar{\tau}_1$, often only one of the solutions, labeled $n_{th}$, will be in a range where $p\left(\mathcal{P}_0|n\right)$ and $p\left(\mathcal{P}_1 |n\right)$ are not negligible. Under these conditions, we can derive a closed form for the function $q_p$ in Eq.(\ref{eq:classicalboundpc}) in the form of:
\begin{equation}
q_b=\frac{1}{2}\Bigg(\text{erf}\Big[\frac{n_{th}-\bar{n}_s \bar{\tau}_0}{\sqrt{2 }\sigma^{(0)}_G}\Big]-\text{erf}\Big[\frac{n_{th}-\bar{n}_s \bar{\tau}_1}{\sqrt{2} \sigma_G^{(1)}} \Big]\Bigg) \label{qg}.
\end{equation}
As indicated before, the expression in Eq.(\ref{qg}) holds for certain regimes, and a more general solution, as well as an explicit expression for $n_{th}$, is reported in the Supplementary Materials.

\subsection{Numerical study}
\begin{figure*}
\includegraphics[width=2\columnwidth]{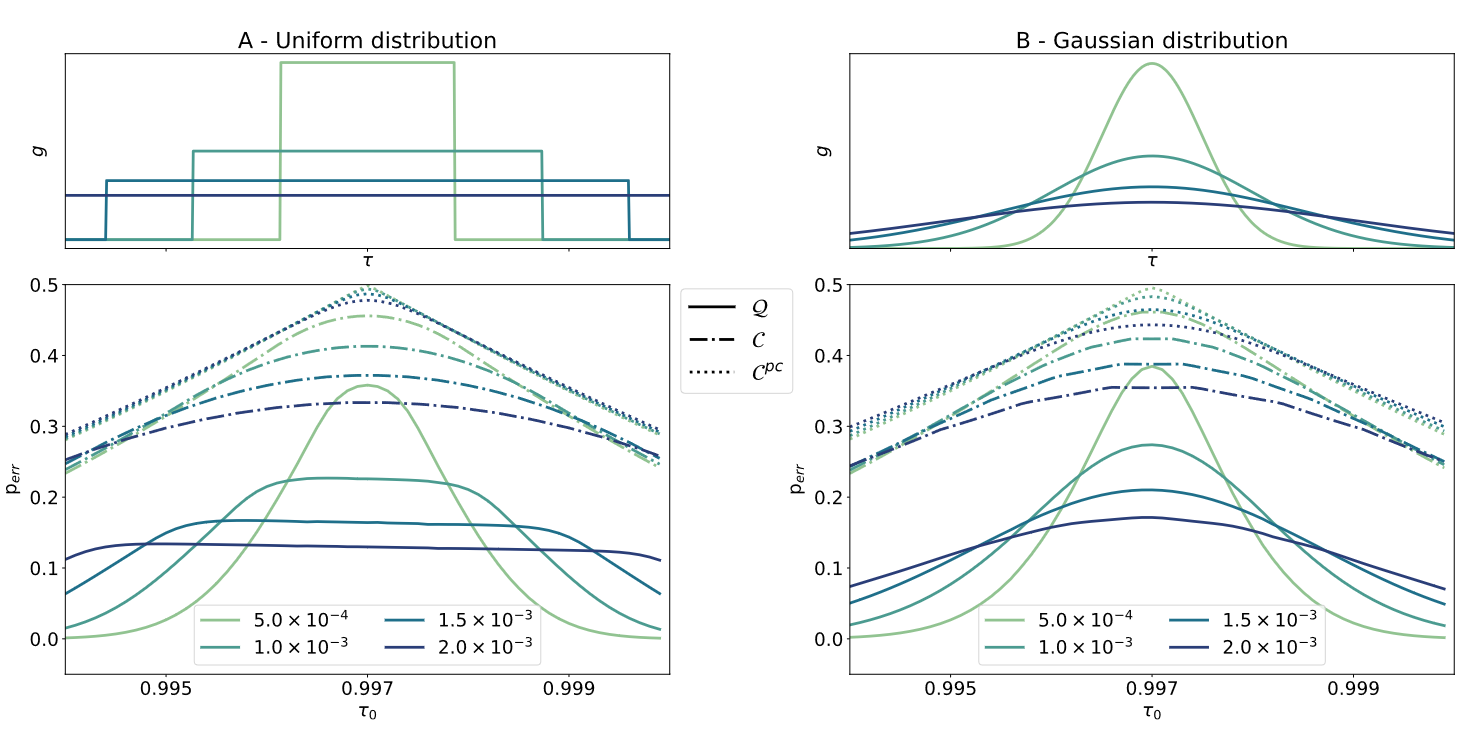}
\caption{\label{fig:bias}\emph{Error probabilities}. The error probabilities as functions of the mean value of the reference process $\bar{\tau_0}$ are displayed, for different variances of the defective process $\sigma^2$ (the numerical values are reported in the legend), and for the different discrimination strategies described in the main text: the quantum strategy (solid line), the classical strategy with PC (dotted line) and the optimal classical strategy (dash-dotted line). The reference distribution is considered strongly peaked: $g_0 \approx \delta\left(\tau-\bar{\tau_0}\right)$. In column A, the defect distribution, $g_1(\tau)$, is uniform, in B it is Gaussian. The distributions $g_1(\tau)$ for different values of variance are plotted on the first row panels. In both bottom panels, the mean number of signal photons is fixed to $n_s=10^5$ and the mean value of the defective process is $\bar{\tau_1}=0.997$.}
\end{figure*}
In the strategies described in the previous sections, the performance of the QCT is a function of the mean photon number $\bar{n}_S$, as well as the form of the distributions of the processes considered, $g_0$ and $g_1$.

A visualization of the discrimination problem is depicted in Fig.(\ref{fig:com}), where the measurement strategy considered at the receiver is photon counting. In Fig.(\ref{fig:com}-\textbf{A}) the distributions of the processes to be discriminated are shown, $g_0$ in red and $g_1$ in blue. In the first row, the two distributions barely overlap. Nonetheless, when a probe with finite energy is used to perform the discrimination, the intrinsic noise of the state on the photon number distribution results in a significant amplification of the overlap of the measurement outcome distributions. This is shown in Fig.(\ref{fig:com}-\textbf{C}), where the two possible photon number distributions, after the interaction with the SUT, and for the classical input state discussed in the previous sections, are plotted. The overlap area, highlighted in green, is a visual representation of the probability of error: $p_{err} $ is equal to half that area. This overlap cannot be reduced using classical states without an increment to the total energy used. However, using the same signal energy, an improvement can be achieved using quantum correlations. If we consider as input a TMSV state, as described earlier in the text, we can analyze the photon number distribution for the signal system, conditioned to having measured a given photon number in the idler branch. This analysis is shown in Fig.(\ref{fig:com}-\textbf{E}), where the overlap between the two distributions is dramatically reduced, leading to much more efficient discrimination. The nature of the advantage resides in the photon number correlations, which are used to greatly reduce the photon number noise of the initial state. In the second row, a situation in which the overlap of the parameter's distributions' is significant is depicted. Indeed, in Fig.(\ref{fig:com}-\textbf{B}, the considered distributions have the same mean value, but different variances. In this case, using a classical discrimination strategy performs very poorly, as shown in Fig.(\ref{fig:com}-\textbf{D}), where the two processes are almost indistinguishable. Fig.(\ref{fig:com}-\textbf{F}) shows how, once again, quantum correlation can be used to greatly improve the performance. We note that the best performances (shown in Fig.(\ref{fig:com}-\textbf{E} and \ref{fig:com}-\textbf{F}) are achieved in the limit of large $K$ so that $\bar{n}_S/K \ll 1$.

From this point on, in order to reduce the number of considered parameters, we assume the reference process $\mathcal{P}_0$ to be strongly peaked around $\tau_0$ so that we can approximate it as $g_0\left(\tau\right) \approx  \delta\left(\tau-\bar{\tau_0}\right)$, where $\delta (\tau)$ is the Dirac delta distribution. We also consider two different noteworthy forms for $g_1$: the Gaussian distribution, $g=G_{\bar{\tau},\sigma}$, and the uniform one, $g=\mathcal{U}_{\bar{\tau},\delta}$. Using a Gaussian probability distribution for the transmittance $\tau$ is justified by the wide range of physical phenomenons it can describe. In this case, the processes are fully characterized by their mean value $\bar{\tau}$ and variance $\sigma^2$. On the other hand, the uniform distribution, $\mathcal{U}_{\bar{\tau},\delta}$ is well suited to describe situations in which there is a complete lack of knowledge of the nature of the process, whose range can be limited by the physical constraints of the apparatus. Once more, the processes are characterized by two parameters only: their mean $\bar{\tau}$ and half-width $\delta$. To make a fair comparison of the error probabilities for Gaussian and uniform distributions, we choose their parameters such that the resulting variances are equal. In particular, we will use $\delta=\sqrt{3}\sigma$, since Var$[\mathcal{U}_{\bar{\tau},\delta}]=\delta^2/3$.

In Fig.(\ref{fig:bias}), we show how, for each of the three strategies studied -- the classical optimal, $\mathcal{C}$, the classical strategy with PC, $\mathcal{C}^{pc}$, and the quantum strategy $\mathcal{Q}$ -- the error probability depends on the reference process' transmittance, $\tau_0$. The number of signal photons is fixed to $n_s=10^5$, as well as the mean value of the transmittance of the defective process $\bar{\tau}_1=0.997$. In column (\textbf{A}), the performances of the different strategies are plotted in the case of a uniform distributed defect, while in column (\textbf{B}) the case of a Gaussian distributed one is reported. In both cases, the results for different values of the variance of the process (denoted by different colors) are shown. In the range studied, it can be seen that, the quantum strategy (solid lines) outperforms both the classical lower bound (dashed lines) and the strategy employing a classical probe and PC (dotted lines). Both classical and quantum strategies have, in fact, a maximum error probability when $\tau_0=\bar{\tau}$. This shared feature does not depend on the noise but on the nature of the problem. Indeed, in this configuration, the overlap between the initial distributions $g$ is maximum and, as a result, the error probability is maximum as well. In both panels, we see how, in this high overlap region, the effect of the probe state's noise is such that, when using a classical strategy, the processes result are completely ($p_{err}=0.5$), or almost completely, indistinguishable, while distinguishability is recovered when quantum states with reduced noise are used.


\subsection{Experimental results}
To validate the presence of a quantum advantage for the conformance test, we experimentally implemented the protocol described in the section ``Quantum Strategy" -- with the substantial difference that we operate with a non-ideal detection efficiency $\eta$. The detection efficiency encompasses different processes: the non-unit quantum efficiency of the detectors, the losses due to the different optical elements, and, in the case of correlated photon sources, the imperfect efficiency affecting the measurement of the correlated photons. These losses cannot be distinguished from those caused by a pure loss channel. Hence, the fit with theoretical curves must be made using the substitution $\tau \rightarrow \eta \tau$. In general, the effect is equivalent to a reduction of the probe energy of the same factor $\tau$. For the quantum strategy however another effect is the reduction of the degree of correlation of the state. Due to this in general it is expected that the quantum advantage will be reduced as $\eta$ becomes lower.  The setup used is thoroughly described in the Materials and Methods section.
\begin{figure}
\includegraphics[width=\columnwidth]{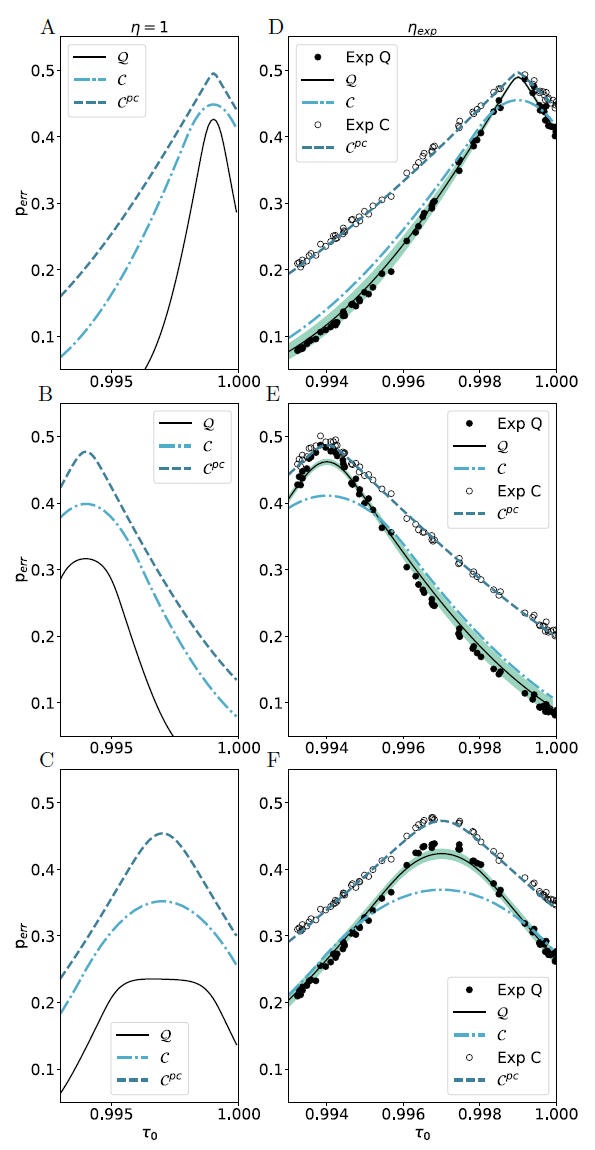}
\caption{\label{fig:explot}\emph{Error probabilities for the QCT}. Error probabilities are plotted as a function of the reference parameter $\tau_0$, for different values of the defective process' mean value $\bar{\tau}$ and variance $\sigma$: $A:\{\sigma = 0.001, \bar{\tau}=0.999\}$, $B:\{\sigma = 0.002, \bar{\tau}=0.994\}$ and $C:\{\sigma = 0.003, \bar{\tau}=0.997\}$. On the left-hand side, we show the theoretical curves (quantum and classical probability of error with PC $\mathcal{Q}$ and $\mathcal{C}^{pc}$, as well as classical bound $\mathcal{C}$), in the ideal case of unitary detection efficiency ($\eta_S=\eta_I=1$) for different sets of parameters $\sigma$ and $\bar{\tau}$. In the right column, we show the experimental error probabilities obtained using the quantum PC protocol (black dots -- Exp Q), and the classical PC one (black circles -- Exp C). The theoretical error probabilities using the estimated experimental parameters (black and light green solid lines - $\mathcal{Q}$ and $\mathcal{C}^{pc}$) are plotted for both protocols, as well as the confidence interval for the quantum case (green shaded areas -- $\pm$ one standard deviation). The classical bound $\mathcal{C}$ is also displayed as a comparison.}
\end{figure}

Our results are presented in Fig.(\ref{fig:explot}). Using a sample presenting a varying spatial transmittance, we acquired data for different values of $\tau$. We used this experimental dataset to realize different defect distributions: in Fig.(\ref{fig:explot}), we plot $p_{err}$ (see Eq.(\ref{eq:1})) when uniform distributions are considered with $A:\{\sigma = 0.001, \bar{\tau}=0.999\}$, $B:\{\sigma = 0.002, \bar{\tau}=0.994\}$ and $C:\{\sigma = 0.003, \bar{\tau}=0.997\}$. For each acquisition, the mean numbers of photons $n_S$ and $n_I$, and the efficiencies $\eta_S$ and $\eta_I$ of the signal and idler channels, are estimated, as well as the electronic noise $\nu$ of the camera. These values are used to draw the theoretical curves: the quantum and classical error probabilities with PC, $\mathcal{Q}$ and $\mathcal{C}^{pc}$, as well as the classical optimal bound $\mathcal{}C$, defined in Eq.\ref{eq:classicalbound}. We display the results in Fig.\ref{fig:explot} (D-E-F). As a comparison, we also plot the ideal case $\eta_S=\eta_I=1$ in Fig.\ref{fig:explot} (A-B-C). For the quantum error probability, we also report the confidence region at two standard deviations as a colored band around the curve. The two sets of experimental data reported in Fig.(\ref{fig:explot}) correspond to the quantum and classical PC strategies respectively. More details on the elaboration of these results and data analysis can be found in the section 'Materials and Methods'.

For the three regimes considered, most points fall within the confidence interval. These results show that, even in the case of degraded detection efficiency, the quantum strategy always brings an advantage with respect to the classical one based on PC. However, within the range of the defect distribution, the optimal classical bound $\mathcal{C}$ on the error probability  becomes smaller than the quantum strategy error probability in some region. To bring this point into perspective, we point to Fig.\ref{fig:explot} (A-B-C), which are constructed with the same experimental parameters but unit detection efficiency. In this case, the quantum strategy overcomes any classical one. We note that, as expected, while the classical error probabilities are little modified by the change in efficiency $\eta$, the quantum one improves significantly in case of $\eta=1$. This effect stems from the fact that, as mentioned, spurious losses $\eta$ reduce the photon number correlations between signal and idler channels.

\subsection*{Constrained probability of error}
\begin{figure*}
\includegraphics[width=2\columnwidth]{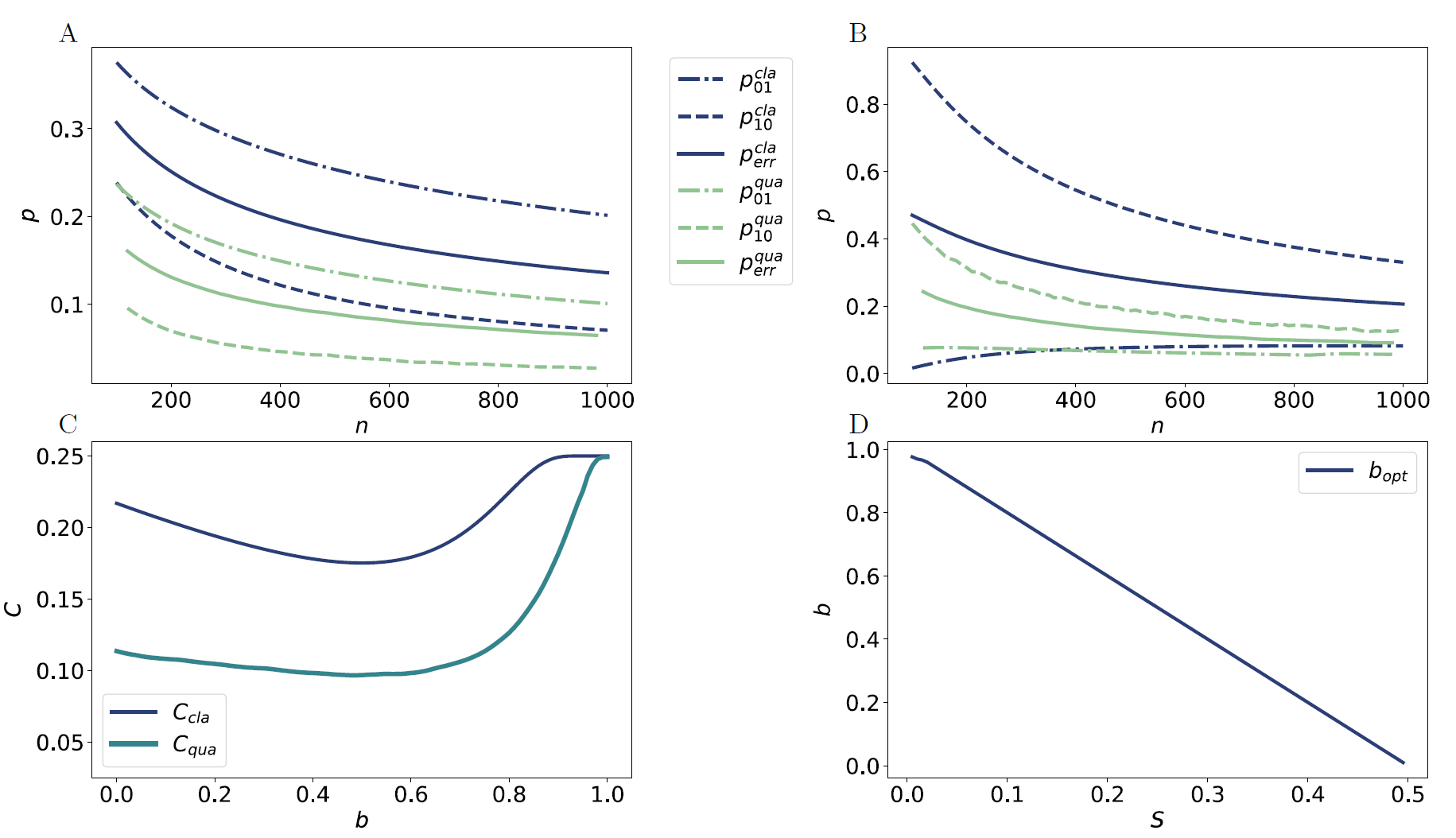}
\caption{\label{fig:fignew} \emph{Cost analysis in QCT}. In panel \textbf{A}, the dependence on the signal photon number $n_S$ of the total error probability ($p_{err}$), false positive ($p_{01}$) and negative ($p_{10}$) is shown. The maximum likelihood post-processing used in this case minimizes $p_{err}$. The reference process is considered strongly peaked around the value $\tau_0=0.8$, while the defective one is chosen uniformly distributed with mean $\bar{\tau}=0.9$ and half-width $\delta=0.09$. In panel \textbf{B}, a situation similar to that of panel \textbf{A} is considered, but we use a biased maximum likelihood post-processing with bias coefficient $b=0.6$ (see main text for details). In panel \textbf{C}, we fix photon number to $n_S=500$ and analyze the dependence of the cost $C$ on $b$. All the other parameters are equal to those of the previous panels. In panel \textbf{D}, the optimum value of $b$ as a function of $S$ is shown.}
\end{figure*}
Up until now, the figure of merit considered was the total probability of error, $p_{err}$ (Eq.\ref{eq:1}). However, this quantity may not be the most relevant one to optimize, depending on the nature of the process under test. In some instances, it can be required to impose a constraint in either one of the conditional probabilities $p_{01}$ or $p_{10}$, instead of their sum. These two types of error represent, in fact, different outcomes and the minimization of one can be deemed more important than that of the other. To proceed further, we use the notion of cost, which quantifies the fact that in the mislabelling of a process, the false-positives and false-negatives may not be equivalent for the operator making the decision: one type of error may be more ``costly" than the other. We introduce the coefficient $0< S< 1$, and define the total cost $C$ of the conformance test as
\begin{equation}
C=S p_{10}+(1-S)p_{01}. \label{cost}
\end{equation}
From Eq.(\ref{cost}), it follows that if we use the cost $C$ as the figure of merit for the evaluation of the QCT, the minimum probability of error is achieved for $S=1/2$. If $S\neq 1/2$, the minimization of the cost will, in general, not minimize the total error probability $p_{err}$.

In the following, we analyze how the strategies presented in the previous sections, consisting of PC and a maximum likelihood decision, both for classical and quantum probes, can be modified to minimize the cost $C$. In the formalism of the previous sections, we write the general term for the probabilities $p_{ij}$ as:

\begin{eqnarray}
p_{01}&=&\sum_{\textbf{n}} p(\textbf{n}|\mathcal{P}_1) \Theta [ p(\textbf{n}|\mathcal{P}_0)-p(\textbf{n}|\mathcal{P}_1)   ]  \nonumber \\
p_{10}&=&\sum_{\textbf{n}} p(\textbf{n}|\mathcal{P}_0) \Theta [p(\textbf{n}|\mathcal{P}_1)-p(\textbf{n}|\mathcal{P}_0)   ] 
\label{conditioned}
\end{eqnarray}
where $\Theta$ is the step function.

The false-positive and false-negative probabilities are plotted as functions of the number of signal photons in Fig.(\ref{fig:fignew}-\textbf{A}) for the previous section's strategies, in the case where $p_{err}$ is the optimized quantity. Both in the classical (red lines) and quantum (blue lines) cases, the false-negative probability tends to be smaller than the false-positive one, meaning that the procedure is more likely to select the reference process. This unbalance comes from the fact that we consider as a reference process a distribution strongly peaked around $\tau_0$, while, for the defective process, we consider a uniform distribution. In terms of photon counts probability, this results in the fact that, when an overlap is present, the most peaked process, $\mathcal{P}_0$, is chosen, while the defective one is selected in a range that is larger, but also less likely to be measured. This results in a bias towards $\mathcal{P}_0$. 

The maximum likelihood post-processing used up until now minimizes the probability of error by construction. To minimize the cost $C$, a different post processing is needed. We can modify the maximum likelihood condition used for the decision, $p\left(\bm{n}|\mathcal{P}_y\right)\geq p\left(\bm{n}|\mathcal{P}_{1-y}\right)$, to make it more likely that one specific process is selected according to some parameters. We select $y$ when:
\begin{equation}
B^{(y)} p\left(\bm{n}|\mathcal{P}_y\right)\geq B^{(1- y)}p\left(\bm{n}|\mathcal{P}_{1- y}\right)
\end{equation} 
where:
\begin{align}
    B^{(0)}&= \frac{1-b}{2} \nonumber \\
    B^{(1)}&= \frac{1+b}{2}
\end{align}
with $b \in [-1,1]$ a real constant.  
We call this new post-processing biased maximum likelihood. Selecting a positive $b$ results in the defective process being chosen more often, while a negative $b$ results in a more likely selection of the reference one. In other words, the value of $b$ can be varied to shift $p_{01}$ and $p_{10}$, reducing the cost function $C$ at the cost of the increase of the total error probability. This is shown in Fig.(\ref{fig:fignew}.\textbf{B}): $p_{01}$ and $p_{10}$ have been brought closer to each other and the total error probability slightly increased, both in the quantum and classical cases. It is worth highlighting how the quantum advantage that we had with maximum likelihood processing is well preserved when the biased maximum likelihood one is used. The optimization of $C$ is performed by varying the coefficient $b$. As an example, we consider the particular situation in which $S=1/4$, meaning that each false positive is considered three times as costly as a false negative. In this situation, it is convenient to use a positive value for $b$, whose effect is to reduce the number of times the reference process $\mathcal{P}_0$ is selected overall, thus reducing the false positive probability $p_{01}$. The dependence of $C$ on $b^{(0)}$ is shown in fig.(\ref{fig:fignew}-\textbf{C}). Both in the quantum and classical cases, as expected, there is a single optimum value for $b$, for which one has a minimum cost. The dependence of the optimum value of $b$ on $S$, with respect to the cost $C$, is shown in Fig.(\ref{fig:fignew}-\textbf{D}). 

\begin{figure}
\includegraphics[width=\columnwidth]{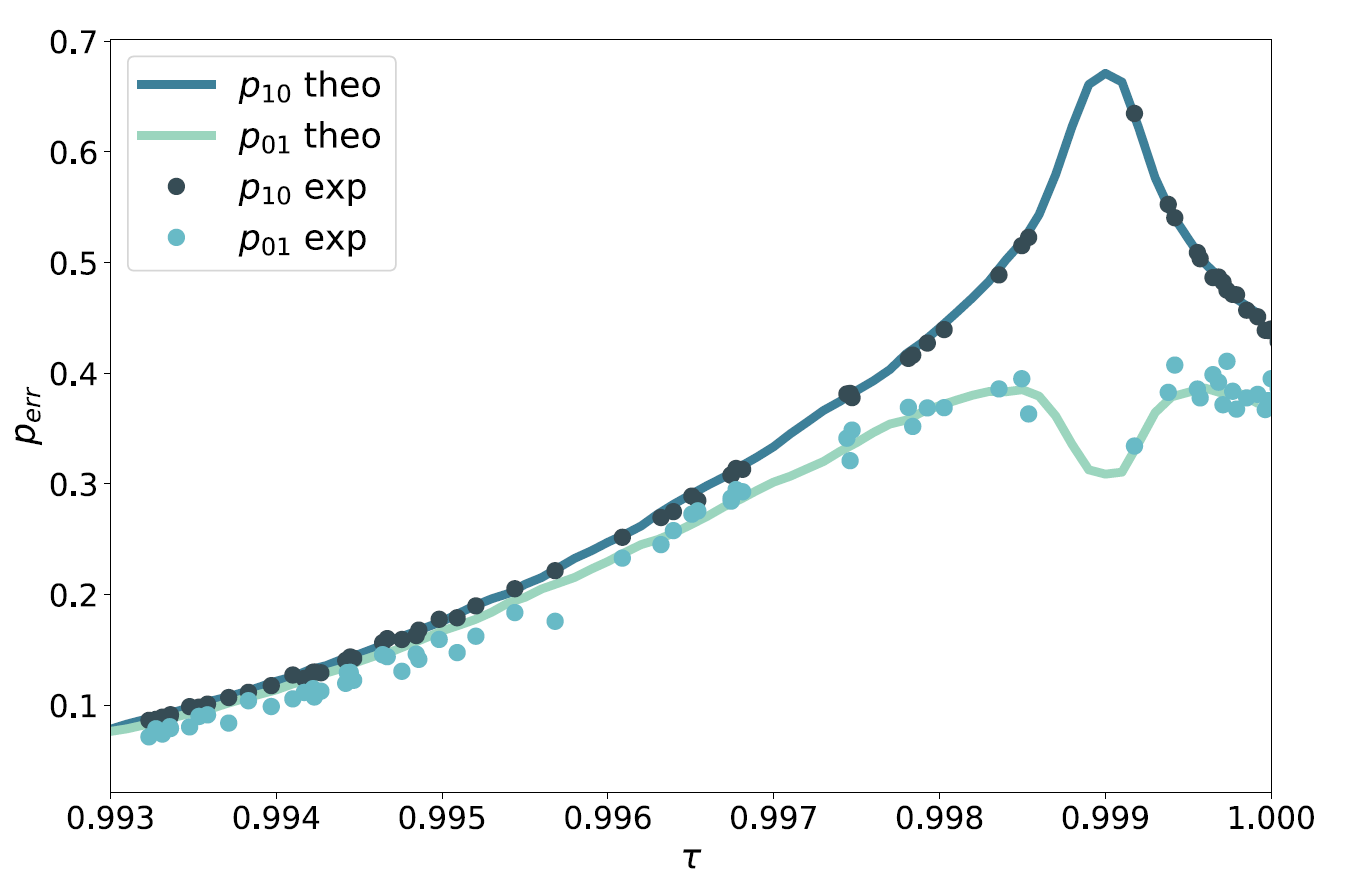}
\caption{\label{fig:expcp}\emph{Theoretical and experimental conditional probabilities of error for the quantum PC strategy:} $\bar{\tau}=0.999$ and $\bar{\sigma}=0.001$.}
\end{figure}

The false positive and negative probabilities of the quantum PC strategy can be evaluated experimentally with the same procedure as the one described in the Materials and Methods for the total probability of error. The experimental points, along with the theoretical curves are reported in Fig.(\ref{fig:expcp}).

\section{Discussion}
In this work, we proposed a protocol addressing the conformance test task, exploiting specific probe states and measurement strategies. We investigated the specific scenario in which the systems under test, representing possible outputs of the unknown process, can be modeled by pure loss channels. 

In this context, we found a lower bound on the error probability that can be achieved in the discrimination task, using any classical source paired with an optimal measurement. We then showed that a particular class of quantum states, namely TMSV states, can be used in conjunction with a simple receiver consisting of a photon-counting measurement and a maximum likelihood decision, to improve the performances over those of any classical strategy. This enhancement is due to the high degree of correlation (entanglement) of TMSV states, whose nature is fundamentally quantum. We also analyzed the particular case of a classical source paired with photon-counting measurements at the receiver.  

We demonstrated that the quantum advantage persists for a wide range of parameters, even in presence of losses due to possible experimental imperfections, and we validated our results by performing an experimental realization of the protocol. We showed an experimental advantage in a realistic scenario where experimental losses amounted to more than $20 \% $,  highlighting the robustness of the proposed protocol.
 
We emphasize the fact that these results are particularly significant because they are achieved using states that are easily produced, as well as a receiver design of simple implementation, allowing practical applications of the protocol with present technology. Remarkably, we found an advantage although the bound $\mathcal{C}$ on the performance of classical states is not tight, meaning that the actual quantum advantage could effectively be higher.
For a thorough study, one can  considered, a scenario in which a set of objects is tested instead of one. In this scenario, we find an informational limit using the Holevo bound and we showed how our proposed quantum strategy, relying on an independent measurement over individual systems, surpasses a more general classical strategy, where joint measurements over a collection of systems are allowed. A detailed description of this result will be presented in a forthcoming paper.
The proposed QCT protocol could be used in the foreseeable future in significant problems concerning the monitoring of production process of any object probed with quantum states. For example, our results on loss channel QCT can be used to boost the accuracy in the identifications of issues in concentration and composition of chemicals production by transmittance measurement.

\section{Materials \& Methods}

\subsection{Experimental setup}
\begin{figure}[h]
\def\svgwidth{\columnwidth}
\centering
	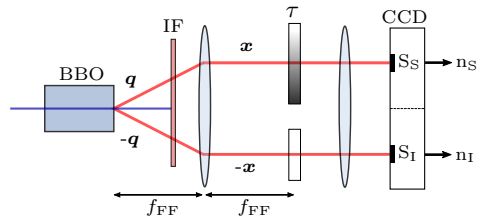
    \caption{\emph{Schematic of the experimental setup}. Pumping a BBO crystal with a laser at $405$ nm, a multi-mode TMSV is generated. Using a lens of focal $f_{\text{FF}}$, the correlation in momentum is converted into correlation in position in the sample plane, which is then imaged on the camera using a second lens. The signal beam passes through the sample of transmittance $\tau$ and is then detected in the $\mathrm{S}_S$ region of the CCD. The idler beam goes directly to $\mathrm{S}_I$, without interacting with the sample, though its optical path is matched with the sample's one using a non-absorbing glass. Integrating the signals over the two detection regions, $n_S$ and $n_I$ are collected.}
	\label{fig:exp2}
\end{figure}

The experiment is based on spontaneous parametric down-conversion and is depicted in Fig.(\ref{fig:exp2}). A multi-mode TMSV state $\otimes^K|\psi\rangle$ is generated using a 1 cm$^3$ type-II $\beta$-barium borate (BBO) crystal and a continuous-wave laser at $\lambda_p=405$ nm, delivering a power of 100 mW. The down-converted photons are correlated in momentum. This correlation is mapped into spatial correlations using a lens in $f$-$f$ configuration: the ``far-field" lens ($f_{\text{FF}} = 1$ cm) is positioned at one focal length of the output plane of the crystal. The absorption sample is positioned in the conjugated plane, which corresponds to the far-field of the source. It consists of a coated glass plate, which presents different transmittance regions, realized with depositions of varying density. A blank coated glass is inserted in the idler beam's path, to match the optical paths of both beams. The sample is imaged using a second ``imaging" lens onto a charge-coupled (CCD) camera (Princeton Instruments PIXIS:400BR eXcellon), working in linear mode, with high quantum efficiency ($>95\%$ at 810 nm) and low (few electrons per pixel per frame) electronic noise. Individual pixels of the camera are binned together in 12x12 macro-pixels, to increase the readout signal-to-noise ratio and the acquisition speed.

The photon counts $n_S$ and $n_I$ of the signal and idler beams are obtained by integrating over two spatially correlated detection areas $\mathrm{S}_S$ and $\mathrm{S}_I$, which are sub-parts of the ``full" correlated areas. The two ``full” correlated areas are schematically defined by the two illuminated regions on the camera, i.e. the spatial extent of the idler and signal beams. Their precise definition is made using a procedure described in \cite{Samantaray_2017}. The total number of spatial modes collected is $K_s \sim 10^3$, and that of temporal mode is $K_t \sim 10^{10}$ (see \cite{Brida_20100} for more details on the estimation). The mean photon number measured in one region $S_i$ is $\bar{n}_i \sim10^5$. Hence, the mean occupation number per mode is very small $\bar{n}_i/(K_s \times K_t) \sim 10^{-8}$. Under these conditions, the multi-thermal marginal photon-number distributions can be well approximated by Poisson distributions \cite{Mandel_1995}. When the absorptive sample is replaced with an equivalent coated glass plate without any deposition, we estimate, for each pair of regions $\mathrm{S}_S$ and $\mathrm{S}_I$, the corresponding detection efficiencies $\eta_S$ and $\eta_I$, by exploiting the correlations of the SPDC process \cite{Meda_2014, Brida:06}. Indeed, the phase matching conditions, the interference filter, the quality of the optical alignment, and the pixels' properties on the camera all introduce small differences in the photon number correlations that need to be accounted for. With the absorptive sample, for each pair of regions, i.e. for all the different values of $\tau$, $N_{\mathcal{D}_{\tau}} = 2 \cdot 10^4$ frames are recorded ($\mathcal{D}_{\tau}$ designates the experimental dataset for transmittance $\tau$).

\subsection{Data analysis}
Using the experimental data, two error probabilities, $p_{01}$ and $p_{10}$, must be evaluated to access to the total error probability:
\begin{equation}
p_{err} = \frac{1}{2}\left(p_{01}+p_{10}\right).
\end{equation} 
In our analysis, the reference process $\mathcal{P}_0$ is strongly peaked around $\tau_0$. Hence, the evaluation of $p_{10}$ with experimental data is rather straightforward. We use the dataset  $\mathcal{D}_{\tau_0}$, composed of $M$ measurement outcome $(n_{s};n_{i})$, when the transmittance $\tau_0$ is inserted in the signal branch, to compute the experimental frequency of error $f_{10}$. In particular, for each data pair  $(n_{s};n_{i})\in \mathcal{D}_{\tau_0}$, a label $y\in\{0,1\}$ is assigned according to the maximum likelihood strategy described in the main text. In the evaluation, experimental parameters, such as the mean photon number of the source and the detection efficiencies are used, estimated in a calibration phase. $f_{10}$ is then defined as the number of wrong decisions, which in this case is simply the number of occurrences of $y=1$ over $M$. $f_{10}$ will converge to the real probability of error $p_{10}$ as the number of experimental points in the dataset $M$ becomes large.
	
On the other hand, the evaluation of $p_{01}$ requires a more careful approach. In fact, one has to construct an ensemble of experimental data $\mathcal{D}_{\mathcal{P}_1}$, whose data are taken with parameters $\tau$ distributed in a way that is representative of the true probability density $g_1 (\tau)$. Experimentally, we acquire $\bar{M}$ photon number pairs for each of the transmittance $\tau_{i},\; i\in\{1,...,L\}$, i.e. $\bar{M}\cdot L$ data points in total. The transmittance values $\tau_{i}$ are selected in an interval $[\tau_{min},\tau_{max}]$. They may not be distributed uniformly in this interval, because of a non perfect experimental control of the position of the absorption layer, and the uncertainty on the estimation of $\tau_i $. Starting from the experimental dataset, and binning the interval $[\tau_{min},\tau_{max}]$ in $K$ equal sub-intervals of size $2l$ ($2l =(\tau_{max}-\tau_{min})/K$), we define the experimental distribution for the transmittance, $g$, as the normalized histogram
\begin{equation}\label{histo}
g(\tau)= \sum_{k=1}^{K}\frac{d_k}{2l L}\Theta_{k}\left(\tau\right),
\end{equation}
where $d_{k}$ is the number of experimental transmittance values $\tau_i$ that fall in the $k$-th bin, so that the relative number of pairs $(n_{s};n_{i})_k$ is $d_{k}\bar{M}$. $\Theta_{k}\left(\tau\right)$ is a step function, equal to unity in the $k$-th bin and 0 elsewhere. In an ideal case, the experimental distribution $g$ in Eq.(\ref{histo}), should be a uniform distribution in the interval $[\tau_{min},\tau_{max}]$, but, in practice, because of the experimental issues mentioned above, it can present significant discrepancies with respect to it.
	
In order to approximate a general probability distribution $f(t)$, we can multiply the coefficients in Eq. (\ref{histo}), by proper weights $w_k$, with $k\in\{1,...,K\}$. Once the weights are determined by a proper optimization procedure, described in the Supplementary Materials, they are used to modulate the number of experimental data in each bin according to the substitution $d_{k}\bar{M}\longmapsto d_{k} w_{k} \bar{M}$. $w_{k}$ represents the fraction of data kept in bin $k$. The ensemble of data randomly picked from the initial set, according to the weights, will define the final set $\mathcal{D}_{\mathcal{P}_1}$. We refer to the distribution of $\mathcal{D}_{\mathcal{P}_1}$ after the reweighting procedure, as $g_{\mathbf{w}}(\tau)$, where $\mathbf{w}=[w_{1},...,w_{K}]$. Since the experimental set cannot be increased, the constraint $0\leq w_k\leq1$ must be imposed. The procedure will in general reduce the number of points in the dataset: an increase of the statistical uncertainty can arise. Thus, it may be useful to introduce a second constraint on the size of the final dataset, $M_{T}=\sum_{k=1}^{K} d_{k} w_{k} \bar{M} $, in the optimization procedure to put a lower bound on how much data is discarded.

For the optimization process, we define the objective function to be maximized as the Bhattacharyya coefficient between the distributions $f$ and $g_{\mathbf{w}}$:
\begin{equation}
T\left(\mathbf{w}\right) = \int_{\tau_{min}}^{\tau_{max}} \sqrt{f(\tau)g_{\mathbf{w}}(\tau)}d\tau.
\end{equation}
This quantity measures their similarity and ranges between 0 and 1. $T\left(\mathbf{w}\right)$ gives a quantitative measure of how close the experimental dataset can be arranged to resemble the objective distribution. We define a threshold value $0\leq T_{th}\leq1$, above which the approximation of $f\approx g_{\mathbf{w}} $  is deemed ``good enough'', i.e. $T\left(\mathbf{w}\right)\geq T_{th}$. In general, the optimal $\mathbf{w}$ and the corresponding  $T\left(\mathbf{w}\right)$, depend on the interval $[\tau_{min};\tau_ {max}]$, which should be chosen so that $\int_{\tau_{min}}^{\tau_{max}}f(t)$ is close to unity. We note that the minimum number of data points $M_{T}$ and the initial distribution of the experimental data can affect the results of the optimization algorithm. For this reason, in our realization of this algorithm, we sampled the interval $[\tau_{min},\tau_{max}]$ with a dense array of values $\tau_i$, taken equispaced within the experimental uncertainty, and we took a very large total number of data points with respect to the target $M_T$. The approximated distribution for the data, reported in the Results section, had a coefficient $T\sim 1$. A more formal description, as well as further details on the procedure, are reported in the Supplementary Materials.
\section*{Acknowledgments}
\subsection*{Funding}
This work was founded by the EU via ``Quantum readout techniques and technologies'' (QUARTET, Grant agreement No 862644). 
\subsection*{Author contributions}
GO and IRB proposed the QCT protocol. GO developed the theoretical model and performed the theoretical and numerical analysis both for the classical limits and the quantum protocol. PB and EL acquired the experimental data, the experiment being designed by IRB. PB and GO performed the data analysis, with contributions from IPD. MG leads the quantum optics group at INRiM and coordinated the project. All authors contributed to the discussion of the results and the writing of the manuscript.
\subsection*{Competing Interests}
The authors declare no competing interest.
\subsection*{Data availability} All data needed to evaluate the conclusions are reported in the paper. Further data are available under reasonable request to the corresponding author.

\bibliography{bib}

\clearpage

\onecolumngrid

{\huge \centerline{\textbf{Supplementary Materials}}}

\vspace{25pt}
\twocolumngrid

\setcounter{figure}{0}
\renewcommand{\thefigure}{S.\arabic{figure}}

\setcounter{equation}{0}
\renewcommand{\theequation}{S.\arabic{equation}}
\section{Derivation of the classical limit}
In this section, we show how the bound $\mathcal{C}$ in Eq.(7) of the main text can be derived. $\mathcal{C}$ is a bound on the minimum probability of error, in the QCT protocol, achievable using as probe classical states -- as defined in the main text --, $\rho_{cla}$, with fixed signal energy $\bar{n}_S$. 

Given an input state $\rho^{cla}$, the output state, after interaction with the SUT, will be either $\rho_0^{cla}$ or $\rho_1^{cla}$. The best performance in discriminating between those states is achieved by using a POVM measurement, which assumes the Helstrom projectors as elements \cite{Helstrom_1976}, yielding a probability of error:
\begin{equation}
p_{err} = \frac{1}{2}\left(1 - D\left(\rho_0, \rho_1\right)\right) \label{hb}
\end{equation}
where $D\left(\rho_0, \rho_1\right) = ||\rho_0 - \rho_1||/2$ is the trace distance and $||\rho||=\sqrt{\rho^{\dagger}\rho}$. A lower bound on the probability of error can be found by upper bounding $D\left(\rho_0, \rho_1\right)$.
\subsection{Strongly peaked reference process}
We consider first a scenario in which only  the defective process $\mathcal{P}_1$ is distributed according to some arbitrary distribution $g_1(\tau)$, while the reference process $\mathcal{P}_0$ is strongly peaked, i.e. it can be assumed that all the object produced have the same trasmittance parameter $\tau_0$.
In this situation, for any classical input state $\rho_{cla}$, as defined in the main text, using the strong convexity of the trace distance, we can write:
\begin{align}
&D(\rho_0,\rho_1)\leq \int d^{\text{\tiny$2M$}} \bm{\alpha}\; d^{\text{\tiny$2L$}} \bm{\beta}\; P(\bm{\alpha},\bm{\beta}) \times \nonumber \\
&\times D(\mathcal{E}_{\tau_0}(|\bm{\alpha}\rangle\langle \bm{\alpha}|)\otimes|\bm{\beta}\rangle\langle \bm{\beta}|,\mathbb{E}_{\mathcal{P}_1}[\mathcal{E}_{\tau}(|\bm{\alpha}\rangle\langle \bm{\alpha}|)]\otimes|\bm{\beta}\rangle\langle \bm{\beta}|) \nonumber \\
&=\int d^{\text{\tiny$2M$}}\bm{\alpha} \, P(\bm{\alpha})D(\mathcal{E}_{\tau_0}(|\bm{\alpha}\rangle\langle \bm{\alpha}|),\mathbb{E}_{\mathcal{P}_1}[\mathcal{E}_{\tau}(|\bm{\alpha}\rangle\langle \bm{\alpha}|)]),
\end{align}
where $\mathbb{E}_{\mathcal{P}_1}$ is the expectation value over the distribution $g_1(\tau)$ and the last equality follows from the fact that the unaffected reference system doesn't change the trace distance. Using once again the convexity of the trace on the second argument of the integrating term $D$, we have:
\begin{align}
&D(\mathcal{E}_{\tau_0}(|\bm{\alpha}\rangle\langle \bm{\alpha}|),\mathbb{E}_{\mathcal{P}_1}[\mathcal{E}_{\tau}(|\bm{\alpha}\rangle\langle \bm{\alpha}|)])  \nonumber \\
&\leq\mathbb{E}_{\mathcal{P}_1}[D(\mathcal{E}_{\tau_0}(|\bm{\alpha}\rangle\langle \bm{\alpha}|),\mathcal{E}_{\tau}(|\bm{\alpha}\rangle\langle \bm{\alpha}|))] \nonumber \\
&= \mathbb{E}_{\mathcal{P}_1}[D(|\sqrt{\tau_0}\bm{\alpha}\rangle\langle \sqrt{\tau_0} \bm{\alpha}|,|\sqrt{\tau}\bm{\alpha}\rangle\langle \sqrt{\tau} \bm{\alpha}|)] \label{con}
\end{align}
To derive this expression, we used the fact that pure loss channels map coherent states into coherent states, i.e. $|\alpha\rangle\xrightarrow{\mathcal{E_{\tau}}}|\sqrt{\tau}\alpha\rangle$. The distance between two pure states is easily calculated:
\begin{align}
D\left(|\sqrt{\tau_0}\bm{\alpha}\rangle,|\sqrt{\tau}\bm{\alpha}\rangle\right))&=\sqrt{1-F(|\sqrt{\tau_0}\bm{\alpha}\rangle,|\sqrt{\tau}\bm{\alpha}\rangle)^2} \nonumber \\
&= \sqrt{1-e^{-|\bm{\alpha}|^2(\sqrt{\tau_0}-\sqrt{\tau})^2}}
\end{align}
with $F(\cdot,\cdot)$ is the fidelity between two quantum states. We can then write:
\begin{equation}
D\left(\rho_0,\rho_1\right)\leq \mathbb{E}_{\mathcal{P}_1}\Bigg[ \int d^{\text{\tiny$2M$}}\bm{\alpha} \, P(\bm{\alpha}) \sqrt{1-e^{-|\bm{\alpha}|^2(\sqrt{\tau_0}-\sqrt{\tau})^2}}\Bigg]
\end{equation}
Given the condition on the total number of photons on the signal system $N_S$  in Eq.(5) of the main text, we have (see Supplementary Materials of Ref.\cite{Pirandola_2011}):
\begin{align}
\int d^{\text{\tiny$2M$}}\bm{\alpha} \, P(\bm{\alpha}) & \sqrt{1-e^{-|\bm{\alpha}|^2(\sqrt{\tau_0}-\sqrt{\tau})^2}}\notag \\
&\leq \sqrt{1-e^{-N(\sqrt{\tau_0}-\sqrt{\tau})^2}}
\end{align}
for any probability distribution $P(\bm{\alpha})$. \\
Using Eq.(\ref{hb}), we can now lower bound the minimum probability of error in the QCT protocol, when one of the processes is strongly peaked, as: 
\begin{equation}
p_{err}^{cla} \geq \frac{1 - \mathbb{E}_{\mathcal{P}_1}\big[\sqrt{1-e^{-\bar{n}_s(\sqrt{\tau_0}-\sqrt{\tau})^2}}\big] }{2} 
\end{equation}
\subsection{Arbitrary reference process distribution}
 If we relax the condition of strongly peaked reference process, so that $\mathcal{P}_0$ is distributed as $g_0(\tau_0)$, we have in place of Eq.(\ref{con}), using the convexity on both arguments:
\begin{align}
&D(\mathbb{E}_{\mathcal{P}_0 }[\mathcal{E}_{\tau}(|\bm{\alpha}\rangle\langle \bm{\alpha}|)],\mathbb{E}_{\mathcal{P}_1}[\mathcal{E}_{\tau}(|\bm{\alpha}\rangle\langle \bm{\alpha}|)]) \leq \nonumber \\
&\mathbb{E}_{\mathcal{P}_0}\mathbb{E}_{\mathcal{P}_1}[D(\mathcal{E}_{\tau_0}(|\bm{\alpha}\rangle\langle \bm{\alpha}|),\mathcal{E}_{\tau_1}(|\bm{\alpha}\rangle\langle \bm{\alpha}|))]= \nonumber \\
&=\mathbb{E}_{\mathcal{P}_0} \mathbb{E}_{\mathcal{P}_1}[D(|\sqrt{\tau_0}\bm{\alpha}\rangle\langle \sqrt{\tau_0} \bm{\alpha}|,|\sqrt{\tau_1}\bm{\alpha}\rangle\langle \sqrt{\tau_1} \bm{\alpha}|)]
\end{align}
So that repeating the steps of previous section the probability of success will be given by
\begin{equation}
p_{err}^{cla}\geq \frac{1 - \mathbb{E}_{\mathcal{P}_0}\mathbb{E}_{\mathcal{P}_1}\big[\sqrt{1-e^{-N(\sqrt{\tau_0}-\sqrt{\tau_1})^2}}\big] }{2},
\end{equation}
That is Eq.(7) of the main text

\section{Classical Limit with photon counting}
In this section, we will derive the best performance that can be achieved using a classical input state, $\rho_{cla}$, paired with a photon counting measurement. In this scenario, idler modes are not required and the minimum probability of error is given by an input state whose signal photon number distribution is Poissonian. The effect of a pure loss channel $\mathcal{E}_\tau$ on the photon number distribution, $P_{\bar{n}_s}(n)$, of a state Poisson distributed with parameter ${\bar{n}_s}$ is to map it into another one having Poisson distribution with parameter ${\bar{n}_s}\tau$: 
\begin{equation}
P_{\bar{n}_s}(n) \xrightarrow{\mathcal{E}_\tau} P_{{\bar{n}_s}\tau}(n) \label{poiss}
\end{equation}

The total probability of error, $p_{err}$, of the procedure can be evaluated as the average over the two possible processes, $\mathcal{P}_0$ and $\mathcal{P}_1$, of the conditioned probabilities of success :
\begin{equation}
p_{err}=\frac{1}{2}(p_{01}+p_{10}) \label{ps}
\end{equation}
where $p_{ij}$ is the probability that the discrimination procedure will give the outcome $i$ conditioned to the SUT being generated by the process $\mathcal{P}_j$.
 
A SUT generated by the process $\mathcal{P}_x$ will have transmission $\tau$ distributed with known probability density function $g_x(\tau)$. The conditioned photon number distribution, $p(n|\mathcal{P}_x)$, can be evaluated by taking the expected value over $g_x(\tau)$ of the distribution conditioned on the specific $\tau$:
\begin{equation}
h^{(x)}(n):=p(n|\mathcal{P}_x)=\mathbb{E}_g [p(n|\tau)]=\int_0^1 P_{{\bar{n}_s}\tau} (n) g_x(\tau) d\tau \label{conditioned_p1}
\end{equation}
The terms in Eq.(\ref{ps}) can be evaluated using the expression in Eq.(\ref{conditioned_p1}) and the fact that:
\begin{equation}
p_{ij}=\sum_{\{n|y=i\}} p(n|\mathcal{P}_j) \label{conditioned}
\end{equation}
where the sum is over all values of $n$ such that the outcome of the decision is $y=i$.

To evaluate the integral in  Eq.(\ref{conditioned_p1}), we notice that $h(n)$ is a compound distribution. We have then:
\begin{align}
\mathbb{E}_h[n]&=\mathbb{E}_g[\mathbb{E}_{P_{{\bar{n}_s}\tau}}(n|\tau)]={\bar{n}_s}\mathbb{E}_g[\tau] \label{ev}\\
var_h[n]&=\mathbb{E}_g[var_{P_{{\bar{n}_s}\tau}}(n|\tau)]+var_g[\mathbb{E}_{P_{{\bar{n}_s}\tau}}(n|\tau)]= \nonumber \\
&={\bar{n}_s}\mathbb{E}_g[\tau] +{\bar{n}_s}^2 var_g (\tau) \label{var}
\end{align} 
In the following we will consider two different cases for the distribution $g(\tau)$, the uniform case and the gaussian one.
\subsection*{Gaussian distribution}
We consider the case in which both the processes distributions have Gaussian form with mean $\bar{\tau}_x$ and variance $\sigma_x^2$, i.e. $g_x(\tau)=G_{\bar{\tau}_x,\sigma_x}(\tau)$. According to Eq.(\ref{ev}-\ref{var}), we have for the final photon number distribution, $h^{(x)}_G(n)$:
\begin{align}
N_G^{(x)}&=\mathbb{E}_h[n]={\bar{n}_s} \bar{\tau}_x \\
\sigma_G^{(x)}&=var_h[n]={\bar{n}_s}\bar{\tau}_x +{\bar{n}_s}^2 \sigma_x^2 
\end{align}
For ${\bar{n}_s}$ large enough, the distribution $h_G(n)$ will converge to a gaussian itself:
\begin{equation}
h^{(x)}_G(n)\sim G(n,N_G,\sigma_G) \qquad\qquad\qquad {\bar{n}_s}>>0\label{approxgauss}
\end{equation}
In this approximation the probability of error of the procedure can be evaluated according to Eq.(\ref{ps}) and (\ref{conditioned}) as:
\begin{align}
p_{err}&= \frac{1}{2}\Bigg(\sum_{\{n|y=1\}} p(n|\mathcal{P}_0)+\sum_{\{n|y=0\}} p(n|\mathcal{P}_1)\Bigg) \nonumber \\
&\approx \frac{1}{2}\Bigg(\int_{\{n|y=1\}} G_{N^{(0)}_G,\sigma^{(0)}_G}(n)+\int_{\{n|y=0\}} G_{N^{(1)}_G,\sigma^{(1)}_G}(n)\Bigg) \label{psgaussian0}
\end{align}
when used the assumption ${\bar{n}_s}>>0$ in Eq.(\ref{approxgauss}) to approximate the the sums with integrals over $n$.

To determine the regions of integration, we impose the condition $G_{N^{(0)}_G,\sigma^{(0)}_G}(n)= G_{N^{(1)}_G,\sigma^{(1)}_G}(n)$. This condition defines up to two threshold values, $n_{th}$, for $n$, whose expression, if $\sigma_G^{(0)}\neq\sigma_G^{(1)}$, is:
\begin{widetext}
\begin{align}
n_{th}^{(-)}&=\frac{N_G^{(0)} (\sigma _G^{(1)})^2- N_G^{(1)}(\sigma _G^{(0)})^2-\sqrt{(\sigma _G^{(0)} \sigma _G^{(1)})^2 \left(2 \left(\sigma_G^{(1)}-\sigma _G^{(0)}\right) \left(\sigma_G^{(1)}+\sigma _G^{(0)}\right) \left(\log \left(\sigma_G^{(1)}\right)-\log \left(\sigma _G^{(0)}\right)\right)+\left(N_G^{(1)}-N_G^{(0)}\right)^2\right)}}{\left(\sigma _G^{(1)}-\sigma_G^{(0)}\right) \left(\sigma _G^{(1)}+\sigma_G^{(0)}\right)} \nonumber \\
   n_{th}^{(+)}&=\frac{N_G^{(0)} (\sigma _G^{(1)})^2- N_G^{(1)}(\sigma _G^{(0)})^2+\sqrt{(\sigma _G^{(0)} \sigma _G^{(1)})^2 \left(2 \left(\sigma_G^{(1)}-\sigma _G^{(0)}\right) \left(\sigma_G^{(1)}+\sigma _G^{(0)}\right) \left(\log \left(\sigma_G^{(1)}\right)-\log \left(\sigma _G^{(0)}\right)\right)+\left(N_G^{(1)}-N_G^{(0)}\right)^2\right)}}{\left(\sigma _G^{(1)}-\sigma_G^{(0)}\right) \left(\sigma _G^{(1)}+\sigma_G^{(0)}\right)}  \nonumber \\
\end{align}
\end{widetext}
 and, if $\sigma_G^{(0)}=\sigma_G^{(1)}$:
\begin{equation}
      n_{th}^{(0)}= \frac{\mu_1+\mu_2}{2}
\end{equation}
We consider the first case, i.e. $\sigma_G^{(0)}\neq\sigma_G^{(1)}$, assuming, without loss of generality, $\sigma_G^{(1)}>\sigma_G^{(0)}$, we have $n_{th}^{(+)}>n_{th}^{(-)}$, and the value $y=1$ will be selected if $n>n_{th}^{(+)}$ or $n<n_{th}^{(-)}$. We have:
\begin{align}
p_{err}\approx \frac{1}{2}\Bigg( &\int_{-\infty}^{n_{th}^{(-)}} G_{N_G^{(0)},\sigma_G^{(0)}}(n)+\int^{n_{th}^{(+)}}_{n_{th}^{(-)}} G_{N_G^{(1)},\sigma_G^{(1)}}(n) +  \nonumber \\
+&\int^{\infty}_{n_{th}^{(+)}} G_{N_G^{(0)},\sigma_G^{(0)}}(n)\Bigg)
\end{align}
when we note how the first integral should run from $0$ rather than $-\infty$ but since we are considering ${\bar{n}_s} >> 0$ we can approximate the exact range of integration with the latter ($n_{th}^{(-)}$ can take also negative values but in that case the integral will not contribute to $p_{err}$). Defining:
\begin{align}
E^x(n_{th})=\text{erf}\Big[\frac{n_{th}-N_G^{(x)}}{\sqrt{2} \sigma_G^{(x)}} \Big]
\end{align}
Where erf$(x)$ is the error function. We have:
\begin{align}
p_{err} &\approx \frac{1}{2}\Big(1-q_G\Big) \\
q_G&:=\frac{1}{2}\Big(E^0(n_{th}^{(+)})+E^1(n_{th}^{(-)})-E^0(n_{th}^{(-)})-E^1(n_{th}^{(+)})\Big)
\end{align}
To simplify the expression above, we note how, if $\sigma_G^{(0)}\approx\sigma_G^{(1)}$ one of the threshold values will be far from the center of both distributions so that it can be neglected and the solution can be considered unique. In particular, we can define:
\begin{align}
q_G^{(+)}({\bar{n}_s},\tau_0,\bar{\tau},\sigma):=&\frac{1}{2}\Bigg(E^0(n_{th}^{(+)})-E^1(n_{th}^{(+)}\Bigg) \\
q_G^{(-)}({\bar{n}_s},\tau_0,\bar{\tau},\sigma):=&\frac{1}{2}\Bigg(E^1(n_{th}^{(-)})-E^0(n_{th}^{(-)})\Bigg) 
\end{align}
and in the approximation of similar variances, we have:

\begin{equation}
q_G \approx \begin{cases}
q^{(+)}_G &\text{if} \hspace{5pt} \bar{\tau}_0\leq\bar{\tau}_1\\
q^{(-)}_G &\text {if} \hspace{5pt} \bar{\tau}_0>\bar{\tau}_1
\end{cases} \label{q}
\end{equation}

The function $q_G$ it's monotonically increasing in the number of photons ${\bar{n}_s}$. The approximation in Eq.(\ref{q}) is the result reported in the main text.
\subsection*{Uniform distribution}

\begin{figure}
\includegraphics[width=0.45 \textwidth]{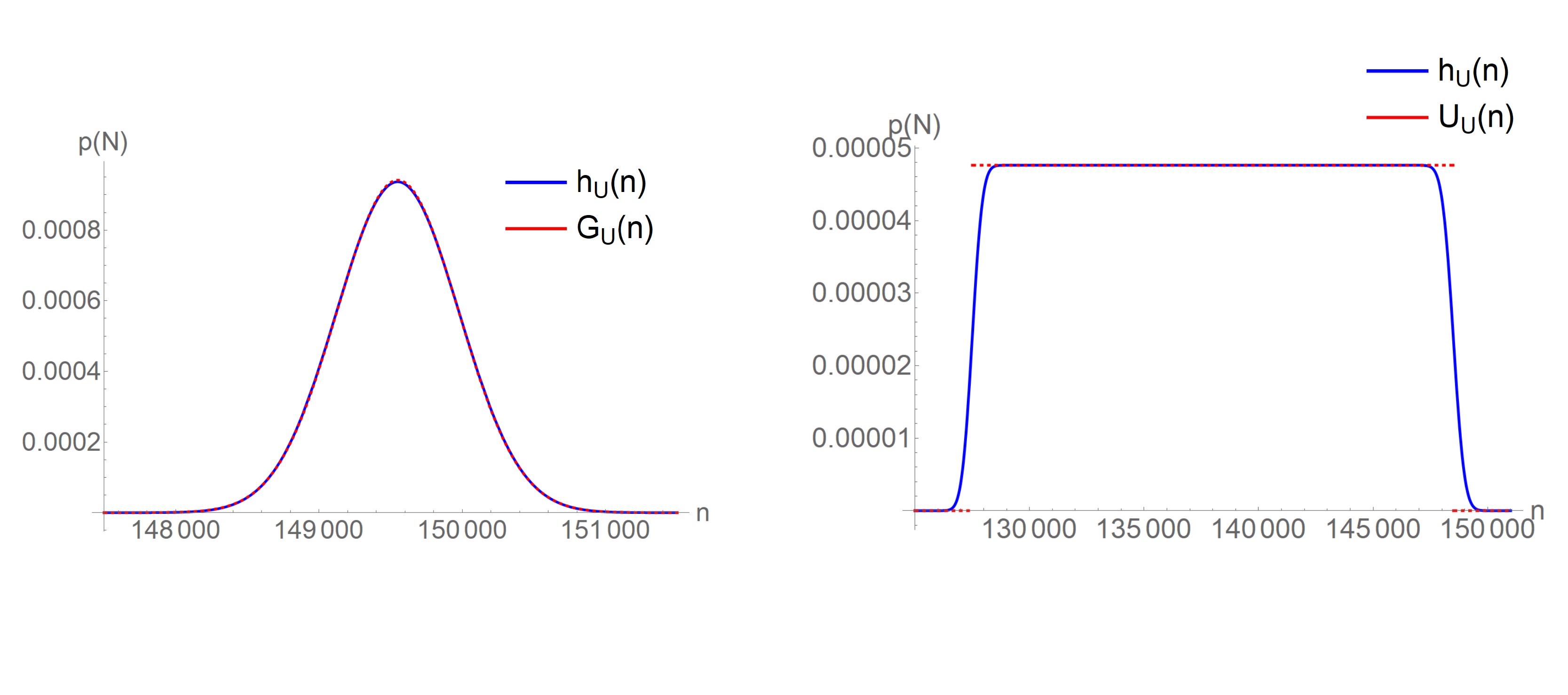}
\caption{\label{fig:apgamma} \textbf{Limiting cases for the distribution $h_U(n)$.} In figure are shown the two limiting cases for the distribution $h_U(n)$, described in the main text. In the first panel the parameters are fixed to ${\bar{n}_s}=1.5*10^5$, $\bar{\tau}$ and $\delta=0.002$, so that $\sigma_U>{\bar{n}_s}\delta$, the situation described in case $(1)$. In the second panel the parameters are ${\bar{n}_s}=1.5*10^5$, $\bar{\tau}=0.92$ and $\delta=0.7$, and we have the situation of case (2). }
\end{figure}

If we consider a uniform distribution having mean $\bar{\tau}$ and half width $\delta$, $\mathcal{U}_{\bar{\tau},\delta}(\tau)$ the integral in Eq.(\ref{conditioned_p1}) can be evaluated analytically. We have:
\begin{align}
h_U(n)=p(n|\mathcal{P}_1)&=\frac{1}{2 \delta}\int_{\bar{\tau}-\delta}^{\bar{\tau}+\delta} P_{{\bar{n}_s}\tau} (n) d\tau= \nonumber \\
 &=\frac{\Gamma(n+1,{\bar{n}_s} (\bar{\tau}-\delta))-\Gamma(n+1,{\bar{n}_s} (\bar{\tau}+\delta))}{2\delta {\bar{n}_s} n! } \label{gamma}
\end{align}
where $\Gamma(a,b)$ is the incomplete gamma function.
Using Eq.(\ref{ev})-(\ref{var}) we have:
\begin{align}
N_U &=\mathbb{E}_h[n]=\bar{\tau}{\bar{n}_s}\\
\sigma^2_U&=var_h[n]=\bar{\tau}{\bar{n}_s}+\frac{1}{3} \delta^2 {\bar{n}_s}^2 
\end{align}  
If we assume, as in the previous section, ${\bar{n}_s}>>0$, the distribution in Eq.(\ref{gamma}) has two noteworthy extreme cases:
\begin{enumerate}
\item
if $\sigma_U>{\bar{n}_s}\delta$, $h_U(n)$ can be approximated with a Gaussian one, $h_U(n)\sim G(n,N_U,\sigma_U)$.
\item 
if $\sigma_U<<{\bar{n}_s}\delta$, $h_U(n)$ can be approximated with a uniform distribution, $h_U(n)\sim \mathcal{U}_{N_U,{\bar{n}_s}\delta}(\tau)$.
\end{enumerate}
The two cases are showed in Fig(\ref{fig:apgamma}).

If we consider case (1), that is the situation we have in our experimental range of parameters,
since we are dealing with a Gaussian distribution the probability of success can be found using the analysis of the previous section. 
\subsection*{Efficiency $\eta$}
In the classical case, inefficiencies resulting in a loss of photons can be accounted for by introducing the efficiency of the channel $\eta<1$. The effect of the efficiency can evaluated using the fact that the action of a loss $\eta$ and the one of the SUT, commute. The final effect of $\eta$ is to reduce the energy available for the discrimination and the classical bounds found can be evaluated accounting for the inefficiency performing the substitution ${\bar{n}_s}\rightarrow\eta{\bar{n}_s}$ in the formulas referring to the perfect efficiency case.

\section{Experimental data analysis}
The probability of error of the QCT procedure, $p_{err}$, can be divided in two separate contributions, namely the probability of error conditioned to the object being produced by the process $\mathcal{P}_0$, $p_{10}$, and   one referring to the process $\mathcal{P}_1$, $p_{01}$:
\begin{equation}
p_{err}=\frac{1}{2} \Big(p_{10}+p_{01}\Big)
\end{equation}
The process $\mathcal{P}_0$ is supposed to be strongly peaked around a value of transmission $\tau_0$, so to evaluate experimentally the contribution $p_{10}$, one can use an object having transmission $\tau_0$ and repeat the discrimination procedure $N_\mathcal{D}$ times. The frequency of error $f_\mathcal{D} (\mathcal{P}_1|\mathcal{P}_0)$ on the dataset $\mathcal{D}$ will converge to the probability of error $p_{10}$ as $N_\mathcal{D}\to \infty$.

The process $\mathcal{P}_1$, on the other hand, will have an arbitrary, but known, probability density $f(\tau)$. To estimate the probability $p_{01}$, a dataset $\mathcal{D}$, representative of the density $f(\tau)$, must be constructed. A convenient way to do this is to create experimentally a collection of $L$ dataset $\mathcal{D}_i$ each generated from a different value of trasmission $\tau_i$, $1 \leq i \leq L,$ in a range $[\tau_{min},\tau_{max}]$ determined by the distribution $f(\tau)$ one wants to approximate.
Suppose for example the target distribution is a uniform one centered in $\bar{\tau}$ and having half-width $\sigma$. To approximate this distribution, measurements are taken with $L$  different transmission $\tau$ equispaced in the interval $[\tau_{min}=\bar{\tau}-\sigma,\tau_{max}=\bar{\tau} +\sigma]$. For a Gaussian distribution, having mean $\bar{\tau}$ and variance $\sigma^2$, measurements can be taken in the interval $[\bar{\tau}-k\sigma,\bar{\tau} +k\sigma]$, where $k$ can be chosen depending on the accuracy required for the approximation. In general, for an arbitrary density $f(\tau)$, one can chose a value $0\leq r \leq 1$ and determine $\tau_{min}$ and $\tau_{max}$ such that:
\begin{equation}
\int_{\tau_{min}}^{\tau_{max}} f(\tau)  d\tau = r \label{limits}
\end{equation}
Each dataset  $\mathcal{D}_i$ will have $N_i$ points. The final dataset used to estimate the gain can be constructed in different ways. The most straightforward one is to take a number of measurements $N_i$, such that the union dataset $\mathcal{D}_U=\bigcup_{i=1}^L \mathcal{D}_i$ is directly distributed  as $f(\tau)$. 

\begin{figure}
\centering
\includegraphics[width=0.45 \textwidth]{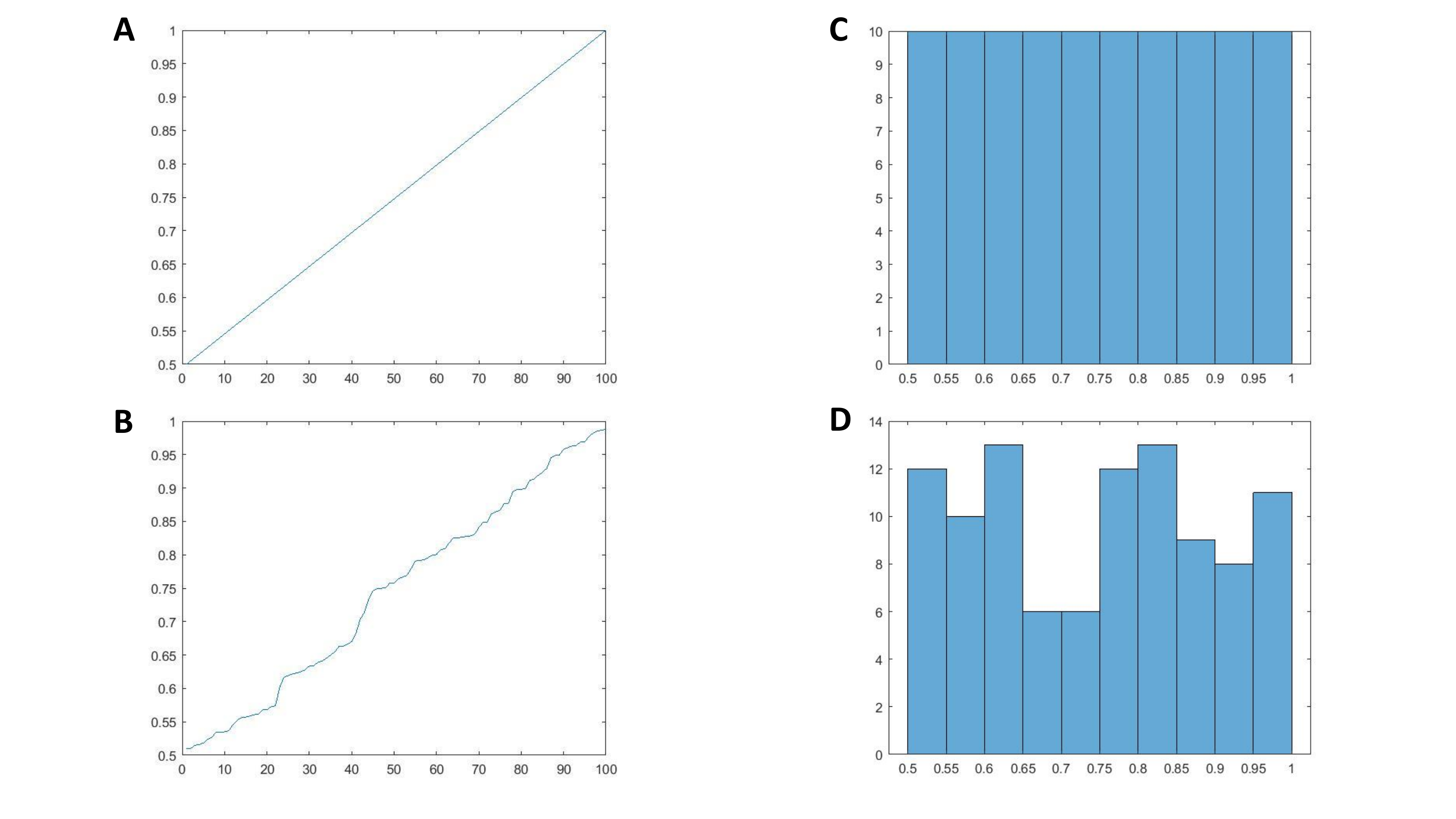}
\caption{\label{fig:random_ideal} \textbf{Sampling of a trasmissivity interval.} In panel \textbf{A} an ideal equispaced sampling for the trasmissivity $\tau$ is reported in the interval $[0.5,1]$. In the $x$ axis are reported the labels $i$(ascending ordered) of the values of trasmissivity $\tau_i$ reported in the $y$ axes. In panel \textbf{B} the same plot is reported in a noisy case, i.e. where the $\tau_i$ are taken equispaced but a value can be assigned up to a certain precision. In panel \textbf{C} and \textbf{D} the histograms referring to the situation of panel \textbf{A} and \textbf{B} respectively are reported.}
\end{figure}

A more flexible approach is to fix the number of measurements taken for each $\tau$ to a given number $\bar{N}$, i.e. $N_i=\bar{N}$ $\forall i$.  
Following this procedure, the union dataset $\mathcal{D}_U$ will be composed of $N_U= L \bar{N}$ measurements extracted from an ideally uniform distribution in $[\tau_{min},\tau_{max}]$. The final dataset can then be rearranged with a procedure of statistical weighting, described in detail in the following, consisting in discarding a certain amount of measurement generated from selected $\tau_i$. The downside of this approach is that the number of measurements to be performed is higher then the number of measurements that will be used in the final dataset approximating $f(\tau)$, $\mathcal{D}_f$. There are however different practical advantages. In practical scenarios in fact the control in selecting an exact value for the transmission $\tau$ is limited, so that it may be difficult to have exactly equispaced values. The situation is showed in Fig.(\ref{fig:random_ideal}).
In panel \textbf{A} and \textbf{B} the plots of an equispaced sampling of the interval $[\tau_{min}=0.5,\tau_{max}=1]$ are reported in the ideal and noisy cases respectively. As the $\tau_i$ are affected by random noise their distribution is affected as well as showed in panels \textbf{C} and \textbf{D}. Panel \textbf{C} refers to the ideal case where the empirical distribution of $\tau$ is uniform as expected. In this case, acting before the measurement only on the coefficients $N_i$ could be an effective way to achieve a good approximation of the target distribution. In the noisy case however, whose distribution is reported in panel \textbf{D}, the distribution is not only non uniform but, in most cases, unknown before the experiment is performed. For this reason, acting only on the coefficient $N_i$, supposing an uniform distribution of the sampling could result in a bad approximation of $f(\tau)$.

The amount of noise varies depending on experimental conditions so in cases in which the deviation from an uniform distribution of the empirical one is small, or in cases in which it can be well characterized beforehand, acting on the coefficient $N_i$ remains a viable solution. Another advantage of the a posteriori statistical weighting procedure is the fact that while some measurements are discarded to construct the final dataset, the procedure can be repeated with the same initial data with a different target distribution, meaning that the gain over different distributions can be evaluated without having to perform the experiment again.

In the following will be given a formal description of the statistical weighting procedure used to create a dataset $\mathcal{D}_f$ approximating one generated by a probability density $f(\tau)$.

\subsection*{Approximation using statistical weighting}
Using the notation of the previous section, the initial experimental dataset $\mathcal{D}_U$ will be composed by  $N_U=L\bar{N}$ measurement results where each group of $\bar{N}$ measurements has been performed on a different transmission $\tau_i$, $0\leq i\leq L$, where the values $\tau_i$ are a sampling of the interval $[\tau_{min},\tau_{max}]$ determined using Eq.(\ref{limits}). 

Our aim is to generate a dataset $\mathcal{D}_f$ to be a good approximation of an ideal dataset generated with measurement with objects distributed as $f(\tau)$. 

To do this let's define the map $\mathcal{H}_\rho^\Pi$, mapping a distribution of $\tau$ onto the distribution $F(\textbf{X})$ of the experimental outcomes $\textbf{X}$ (e.g. the distribution of photon pairs):
\begin{equation}
F(\textbf{X})=\mathcal{H}_\rho^\Pi[f(\tau)] \label{hx}
\end{equation}
The subscript $\rho$ and superscript $\Pi$ denote the dependence of the map on the initial state used to probe $\tau$ and the measurement performed and will be omitted from here on forward.

Our aim is to operate a transformation $W$ on our initial dataset, $\mathcal{D}_U \xrightarrow{W} \mathcal{D}_f$, such that the point in $\mathcal{D}_f$ are distributed as $h(\textbf{X})$, i.e. $\textbf{X}_f=\textbf{X}\in\mathcal{D}_f\sim F(\textbf{X})$. Let us introduce the weight vector $\textbf{w}=[w_1,...,w_L]$. The initial dataset $\mathcal{D}_U$ is composed of $L$ subsets $\mathcal{D}_i$, each composed of $\bar{N}$ experimental points, $\mathcal{D}_i=[\textbf{X}_i^1,...,\textbf{X}_i^{\bar{N}}]$. We define the transformation $W_i$ as the one acting on each $\mathcal{D}_i$, $\mathcal{D}_i \xrightarrow{W_i} \mathcal{D}'_i$, extracting at random a fraction $N'_i=\lfloor w_i \bar{N} \rfloor$ of experimental points from $\mathcal{D}_i$. The final dataset will be composed as the union set of all the $\mathcal{D}'_i$ and the transformation W can be defined as:
\begin{equation}
\mathcal{D}_U \xrightarrow{W} \mathcal{D}_f=\bigcup_{i=1}^L \mathcal{D'}_i
\end{equation}
Therefore a successful approximation is dependent on the optimization of the vector \textbf{w}. 

To perform this optimization, we have to introduce a figure of merit. To do so let's define a probability density function,  $g(\tau)$, approximating the empirical distribution in $\mathcal{D}_U$. This can be done using a normalized histogram. We fist divide the interval $[\tau_{max},\tau_{min}]$ in $K$ sub intervals with equal width, where $K$ is selected via an adequate algorithm. We can define the categories vector  $\textbf{$\tau$}^C=[\tau_1^C,...,\tau_K^C]$, taking $\tau_i^C$ as the middle point between the edges of the $i$-th subinterval. 
The empirical distribution $g(\tau)$ can then be defined as:
\begin{equation}
g(\tau)=\sum_{i=1}^K \frac{c_i}{2l L} \Theta_i (\tau) \label{gtau}
\end{equation}  
where $c_i$ are the number of occurences on the $i$-th sub interval, $L$ is the number of different values of $\tau$ and $\Theta_i(\tau)$ is a step function and $l$ is the half width of each sub interval:
\begin{align}
\Theta_i (\tau)& =  \begin{cases}
1 & \tau_i-l \leq \tau \leq \tau_i + l \\
0 & \text{otherwise}
\end{cases} \\
l& = \frac{\tau_{max}-\tau_{min}}{2 K}
\end{align}
The probability distribution in Eq.(\ref{gtau}) defines a distribution for the experimental data, $G(\textbf{X})$ according to Eq.(\ref{hx}):
\begin{equation}
G(\textbf{X})=\mathcal{H}[g(\tau)] 
\end{equation}
Since the map $\mathcal{H}$ does not depend on the probability distribution we can act on the applying a transformation on the distribution of the dataset we can apply a transformation of the distribution  $g(\tau)$:
\begin{align}
g(\tau)\xrightarrow{W} g'(\tau) & \longrightarrow \mathcal{H}[g(\tau)]=G(\textbf{X})\rightarrow \mathcal{H}[g'(\tau)]= \nonumber \\
&=G'(\textbf{X})\sim F(\textbf{X})=\mathcal{H}[f(t)]
\end{align} 
The transformation $W$ must then act on $g(\tau)$, transforming it in $g'(\tau)\sim f(\tau)$. 
We introduce than the parameterized function $g_\textbf{w} (\tau)$:
\begin{equation}
g_\textbf{w} (\tau)=\sum_{i=1}^K \frac{c_i w_i}{2l L} \Theta_i (\tau)
\end{equation} 
depending on the weight vector $\textbf{w}=[w_1,...,w_K]$ under the normalization constraint:
\begin{equation}
\int g_\textbf{w} (\tau) d\tau=1 \longrightarrow \sum_{i=1}^K \frac{c_i w_i }{L}=1 \label{nor}
\end{equation}
Note how in this configuration, the transformation $W$ acts separately on each of the $K$ sub interval rather than on each of the $L$ values of $\tau$ measured, so that the vector $\textbf{w}$ has dimensionalty $K \leq L$.

It is clear that the initial distribution is $g(\tau)=g_{\textbf{w}_0}$, where $\textbf{w}_0=[1,...,1]$ and the target function is $g'(\tau)=g_\textbf{w'}$, where $\textbf{w'}$ is the optimal weight such that we have the best approximation $g'(\tau)\sim f(\tau)$. In a formal way, we can define the objective function to be maximized:
\begin{equation}
T(\textbf{w})=\int \sqrt{f(\tau) g_\textbf{w}(\tau)} dt \label{BC}
\end{equation}
and the optimal vector $\textbf{w'}$:
\begin{equation}
\textbf{w'}=\text{argmax}_\textbf{w} T(\textbf{w})
\end{equation}
$T(\textbf{w})$ is the Bhattacharyya coefficient (BC) between the distribution  $g_\textbf{w}(\tau)$ and $f(\tau)$, a measure of their similarity, and ranges between $0$ and $1$. 
$T(\textbf{w}')$ gives a quantitative measure of how close our experimental dataset can be arranged to one produced by the distribution $f(\tau)$. We can define a threshold value $0\leq T_{th}\leq 1$ such that  
we accept the approximation if $T(\textbf{w}')\geq T_{th}$ and reject it otherwise.
Before performing the optimization, some other constraint must be imposed. 

The statistical weight represents the fraction of data that we select from each dataset, and considering that the data are considered fixed, i.e. new data cannot be added after the initial dataset has been created, we must impose for the vector $\textbf{w}$ the condition:
\begin{equation}
0\leq w_i\leq 1 \qquad \forall i \label{fdata}
\end{equation}
where the first inequality follows from the fact that the statistical weight has to be non negative.
The fixed data condition in Eq.(\ref{fdata}) is not compatible with the normalization constraint in Eq.(\ref{nor}), but this issue is easily solved taking into account the number of experimental points in each dataset in the formulation of our problem.

The number of experimental data in the original union dataset is $N_\mathcal{U}=L\bar{N}$. Since the transformation $W$ consists exclusively in discarding data, the number of points in the final dataset will be $N_T\leq N_\mathcal{U}$. It is clear that $N_T$ will influence the variance of the estimated gain, so it makes sense to fix it before the optimization process rather then leaving it as a free parameter.
We can then redefine $g_\textbf{w}(\tau)$ as:
\begin{equation}
g_\textbf{w} (\tau)=\sum_{i=1}^K \frac{N_\mathcal{U}}{N_T} \frac{c_i w_i}{2l L} \Theta_i (\tau)
\end{equation} 
The new normalization condition is then:
\begin{equation}
\sum_{i=1}^K \frac{N_\mathcal{U}}{N_T} \frac{c_i w_i }{L}=1 \label{nor1}
\end{equation}
The new definition for $g_\textbf{w}(\tau)$ shifts the normalization condition in Eq.($\ref{nor}$), imposed on the number $L$ of different $\tau$, to the new one in Eq.(\ref{nor1}) where it is imposed on the number of experimental data in the final dataset. With this new definition, the initial distribution will be $g(\tau)=g_{\textbf{w}_0}(\tau)$, where in this case $\textbf{w}_0=[w_0,...,w_0]$ and $w_0=N_T/N_\mathcal{U}$. In this formulation, $\textbf{w}$ is exactly a statistical weight vector, i.e. defines the fraction of experimental data taken from each sub interval dataset, and we can find the optimal vector $\textbf{w}'$, maximizing $T(\textbf{w})$ under the fixed data constraint of Eq.(\ref{fdata}) and the normalization one in Eq.(\ref{nor1}), that are now compatible conditions.
In general, the optimization of eq.(\ref{BC}) under the $K+1$ constraint described can be solved by numerical methods. This procedure will yield a value $\textbf{w}'$ and its value $T(\textbf{w}')$.
It is clear that the choice of sub interval number $K$ influences the outcome of the procedure and could be included in the optimization process. We decided however to non include it to avoid unnecessary numerical complications, as good approximation can be reached in a wide variety of cases performing the binning before the optimization, using well known algorithms. In our case the binning was performed using Sturges rule.

\begin{figure*}
\centering
\includegraphics[width=1 \textwidth]{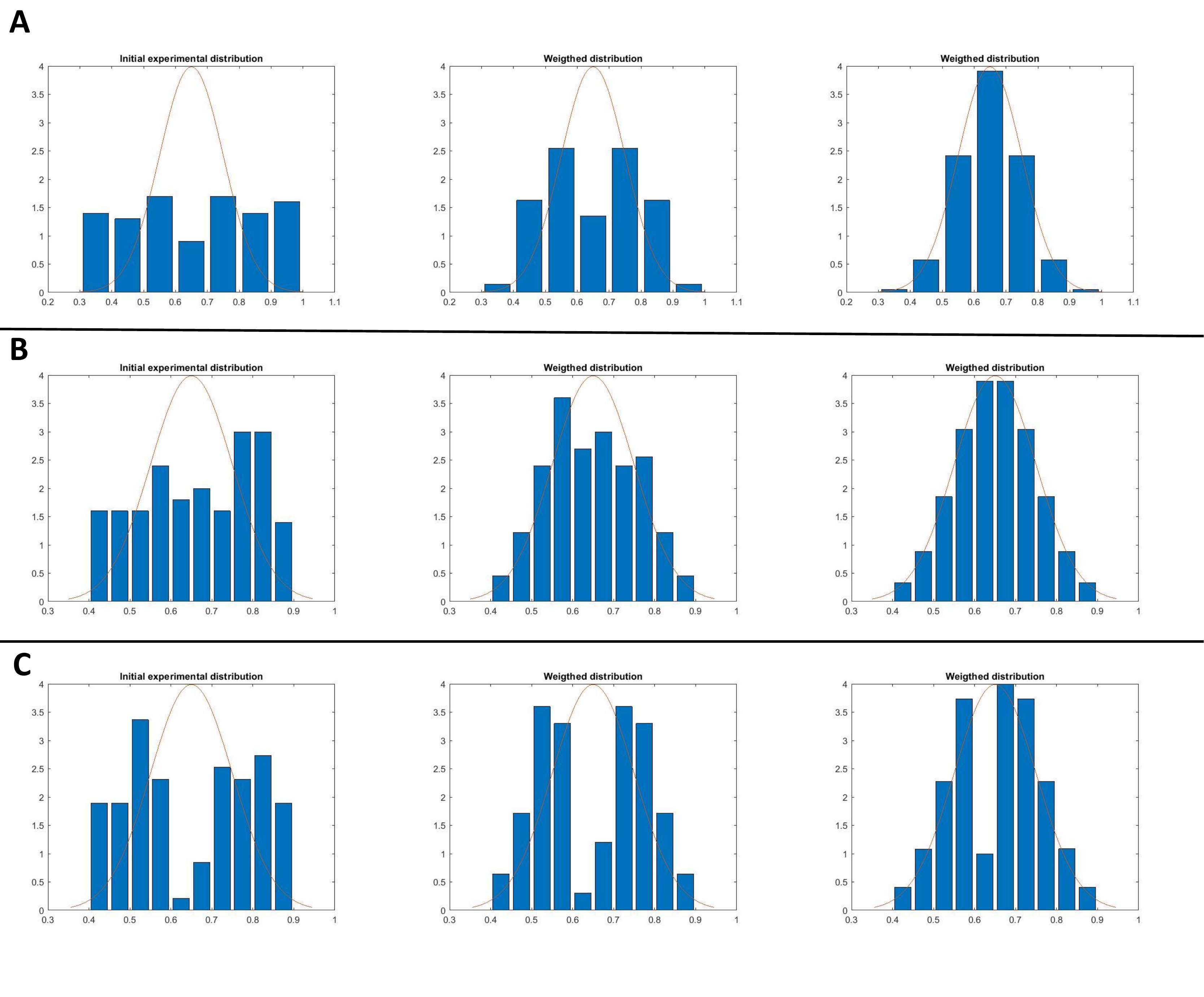}
\caption{\label{fig:cmp_dst} \textbf{Distribution approximation with different initial dataset.} In row \textbf{A} are showed the initial dataset and the best approximation for different values, $\bar{N}=150$ and $\bar{N}=500$, of experimental data taken (see main text for more details). In row \textbf{B} the same situation is reported in case of an initial dataset taken in a smaller interval $[\tau_{min},\tau_{max}]$ w.r.t. row \textbf{A}. In row \textbf{C} is considered a faulty initial dataset, having few measurements around the peak of the target distribution.}
\end{figure*}  

In Fig.(\ref{fig:cmp_dst}) are reported some examples of approximation of a distribution from simulated experimental dataset. The distribution to be approximated is a Gaussian one with mean $\bar{\tau}=0.65$ and standard deviation $\sigma=0.1$ and is reported in each panel as a red curve. In the first row, Fig.(\ref{fig:cmp_dst}).\textbf{A}, the experimental values of $\tau$ are taken from an uniform distribution in $[\bar{\tau}-3.5 \sigma,\bar{\tau}+3.5 \sigma]$, meaning that $\tau_{min}$ and $\tau_{max}$ are selected imposing $r= 0.9995$ in Eq.(\ref{limits}). Starting from the left, the first panel represent the initial experimental distribution. The second and third panels show the best approximation in the situation in which are considered measurement with $L=100$ different values of $\tau$ and, for each $\tau$, the experimental points taken are $\bar{N}=150$(second) and $\bar{N}=500$(third). In both cases, the number of point of the final dataset is fixed to $N_T=10000$. As can be seen, if not enough point can be discarded, as it is the case in the second panel, a good approximation is difficult to reach. When a lot of experimental points are taken, as in the situation of the third panel, as can be expected, a good approximation of most the distributions can be reached. In quantitative terms, we have for the second panel a value of $T^{\bar{N}=150}(\textbf{w}')\approx 0.9$ while for the third $T^{\bar{N}=500}(\textbf{w}')\approx 0.99$. In general, the BC tends to have high values in the situation we are considering, for reference consider that the unaltered initial distribution has a value $T^{0}(\textbf{w}')\approx 0.8$. 
In Fig.(\ref{fig:cmp_dst}).\textbf{B} are reported the same scenario considered in row \textbf{A} but considering an initial dataset taken in a smaller interval, such that $r=0.997$ (half width of $[\tau_{min},\tau_{max}]$ equal $3\sigma$). While a direct comparison may be unfair given that the dataset are randomly generated, it can seen that reducing the interval may be useful since more data will be taken at the peak of the target distribution. In the second panel, referring again to the situation of $\bar{N}=150$ we can see a better approximation with respect to the same panel in row \textbf{A}.  Finally, in row \textbf{C} the same situation is reported in case of a faulty initial dataset, i.e. one having very few values around the peak of the target distribution. In this case even with $\bar{N}=500$ experimental data for each $\tau$ (third panel) a good approximation cannot be reached, indicating that in those situation it may be worth to perform the measurements again.


\end{document}

%% file: exp1.pdf_tex
\begingroup%
  \makeatletter%
  \providecommand\color[2][]{%
    \errmessage{(Inkscape) Color is used for the text in Inkscape, but the package 'color.sty' is not loaded}%
    \renewcommand\color[2][]{}%
  }%
  \providecommand\transparent[1]{%
    \errmessage{(Inkscape) Transparency is used (non-zero) for the text in Inkscape, but the package 'transparent.sty' is not loaded}%
    \renewcommand\transparent[1]{}%
  }%
  \providecommand\rotatebox[2]{#2}%
  \ifx\svgwidth\undefined%
    \setlength{\unitlength}{428.37562024bp}%
    \ifx\svgscale\undefined%
      \relax%
    \else%
      \setlength{\unitlength}{\unitlength * \real{\svgscale}}%
    \fi%
  \else%
    \setlength{\unitlength}{\svgwidth}%
  \fi%
  \global\let\svgwidth\undefined%
  \global\let\svgscale\undefined%
  \makeatother%
  \begin{picture}(1,0.27286029)%
    \put(0.21640046,0.2451405){\color[rgb]{0,0,0}\makebox(0,0)[b]{\smash{{\scriptsize Transmitter}}}}%
    \put(0.76461294,0.24513775){\color[rgb]{0,0,0}\makebox(0,0)[b]{\smash{{\scriptsize Receiver}}}}%
    \put(0.35863121,0.20335269){\color[rgb]{0,0,0}\makebox(0,0)[lb]{\smash{{\scriptsize M (signal)}}}}%
    \put(0.35714442,0.1025277){\color[rgb]{0,0,0}\makebox(0,0)[lb]{\smash{{\scriptsize L (idler)}}}}%
    \put(0,0){\includegraphics[width=\unitlength,page=1]{exp1.pdf}}%
    \put(0.55472975,0.2457274){\color[rgb]{0,0,0}\makebox(0,0)[lb]{\smash{$\mathcal{E}_{\tau}$}}}%
    \put(0.55583144,0.00582415){\color[rgb]{0,0,0}\makebox(0,0)[lb]{\smash{$\mathcal{I}$}}}%
    \put(0.22120327,0.10681362){\color[rgb]{0,0,0}\makebox(0,0)[b]{\smash{$\rho$}}}%
    \put(0.76895729,0.10681584){\color[rgb]{0,0,0}\makebox(0,0)[b]{\smash{$\Pi$}}}%
    \put(0.21975581,0.16793138){\color[rgb]{0,0,0}\makebox(0,0)[b]{\smash{{\scriptsize input state}}}}%
    \put(0.76363797,0.1679283){\color[rgb]{0,0,0}\makebox(0,0)[b]{\smash{{\scriptsize POVM}}}}%
    \put(0,0){\includegraphics[width=\unitlength,page=2]{exp1.pdf}}%
    \put(0.96692159,0.12767188){\color[rgb]{0,0,0}\makebox(0,0)[b]{\smash{DP}}}%
  \end{picture}%
\endgroup%

%% file: exp2.pdf_tex
\begingroup%
  \makeatletter%
  \providecommand\color[2][]{%
    \errmessage{(Inkscape) Color is used for the text in Inkscape, but the package 'color.sty' is not loaded}%
    \renewcommand\color[2][]{}%
  }%
  \providecommand\transparent[1]{%
    \errmessage{(Inkscape) Transparency is used (non-zero) for the text in Inkscape, but the package 'transparent.sty' is not loaded}%
    \renewcommand\transparent[1]{}%
  }%
  \providecommand\rotatebox[2]{#2}%
  \ifx\svgwidth\undefined%
    \setlength{\unitlength}{211.85365241bp}%
    \ifx\svgscale\undefined%
      \relax%
    \else%
      \setlength{\unitlength}{\unitlength * \real{\svgscale}}%
    \fi%
  \else%
    \setlength{\unitlength}{\svgwidth}%
  \fi%
  \global\let\svgwidth\undefined%
  \global\let\svgscale\undefined%
  \makeatother%
  \begin{picture}(1,0.32255121)%
    \put(0.2086287,0.00511366){\color[rgb]{0,0,0}\makebox(0,0)[lb]{\smash{{\scriptsize $f_\textrm{FF}$}}}}%
    \put(0,0){\includegraphics[width=\unitlength,page=1]{exp2.pdf}}%
    \put(0.42371314,0.3060774){\color[rgb]{0,0,0}\makebox(0,0)[lb]{\smash{$\tau$}}}%
    \put(0,0){\includegraphics[width=\unitlength,page=2]{exp2.pdf}}%
    \put(0.34936929,0.00511366){\color[rgb]{0,0,0}\makebox(0,0)[lb]{\smash{{\scriptsize $f_\textrm{FF}$}}}}%
    \put(0.35290952,0.25297453){\color[rgb]{0,0,0}\makebox(0,0)[lb]{\smash{{\scriptsize \boldmath$x$}}}}%
    \put(0.34595564,0.06749441){\color[rgb]{0,0,0}\makebox(0,0)[lb]{\smash{{\scriptsize -\boldmath$x$}}}}%
    \put(0.16932593,0.10991319){\color[rgb]{0,0,0}\makebox(0,0)[lb]{\smash{{\scriptsize -\boldmath$q$}}}}%
    \put(0.17640639,0.2019578){\color[rgb]{0,0,0}\makebox(0,0)[lb]{\smash{{\scriptsize \boldmath$q$}}}}%
    \put(0,0){\includegraphics[width=\unitlength,page=3]{exp2.pdf}}%
    \put(0.68493306,0.2283197){\color[rgb]{0,0,0}\makebox(0,0)[lb]{\smash{{\scriptsize n$_\textrm{S}$}}}}%
    \put(0.68492775,0.08633314){\color[rgb]{0,0,0}\makebox(0,0)[lb]{\smash{{\scriptsize n$_\textrm{I}$}}}}%
    \put(0.07494219,0.20903807){\color[rgb]{0,0,0}\makebox(0,0)[lb]{\smash{{\scriptsize BBO}}}}%
    \put(0.23481924,0.2822438){\color[rgb]{0,0,0}\makebox(0,0)[lb]{\smash{{\scriptsize IF}}}}%
    \put(0.57113658,0.29640441){\color[rgb]{0,0,0}\makebox(0,0)[lb]{\smash{{\scriptsize CCD}}}}%
    \put(0,0){\includegraphics[width=\unitlength,page=4]{exp2.pdf}}%
    \put(0.59637065,0.22913915){\color[rgb]{0,0,0}\makebox(0,0)[lb]{\smash{{\scriptsize S$_\textrm{S}$}}}}%
    \put(0.5966869,0.08892273){\color[rgb]{0,0,0}\makebox(0,0)[lb]{\smash{{\scriptsize S$_\textrm{I}$}}}}%
  \end{picture}%
\endgroup%